\documentclass[prd,showpacs,nofootinbib,preprint, 10 pt]{revtex4}
\usepackage{amssymb}
\usepackage{amsmath,graphicx,color,epsfig}
\usepackage{pstricks}
\usepackage{float}

\begin{document}

\title{Semileptonic $B$ to Scalar meson Decays in the Standard Model with
Fourth Generation}
\author{M. Jamil Aslam}
\email{jamil@ncp.edu.pk; jamil@phys.qau.edu.pk}
\affiliation{Physics Department, Quaid-i-Azam University, Islamabad, Pakistan}
\date{\today }

\begin{abstract}
We study the effects of the fourth generation of quarks on the total
branching ratio and the lepton polarizations in $\bar{B}_{0}\rightarrow
K_{0}^{\ast }(1430)l^{+}l^{-}$( $l$ = $\mu $, $\tau $) decay. Taking fourth
generation quark mass $m_{t{\prime }}$ of about $400$ to $600$ GeV with the
mixing angle $\left\vert V_{t^{\prime }b}^{\ast }V_{t^{\prime }s}\right\vert
$ in the range $(0.05-1.4)\times 10^{-2}$ and using the phase to be $80^{o}$%
, it is found that the branching ratio and lepton polarizations are quite
sensitive to these fourth generation parameters. In future the experimental
study of this decay will give us an opportunity to study new physics
effects, precisely, to search for the fourth generation of quarks $%
(t^{\prime },b^{\prime })$ in an indirect way.
\end{abstract}

\maketitle


\section{Introduction}

The CP violation through the CKM paradigm \cite{Cabibbo, Kobayashi} in the
SM has been extremely successful in explaining most of the experimental
data. However in the past few year a lot more data were accumulated from the
two B-factories and also the improvement in the accuracy of some of the
theoretical calculations led us to understand that several of the
experimental results ar difficult to explain with in the SM\ with three
generations (SM3) \cite{Lunghi, Bona, Lunghi1}. This leads us the think
about some beyond the SM3 scenarios and among them the simplest one is the
Standard Model with fourth generation (SM4). In this model the SM is
enlarged by a complete sequential 4th family of quarks and leptons: a new $%
(t^{\prime },b^{\prime })$ and $(\nu ^{\prime },l^{\prime })$ which are the
heavy chiral doublets. Review and the summary statements of SM4 can be found
in \cite{Frampton}.

During the last years, a number of analysis were published with the goal of
investigating the impact of the existence of a fourth generation on Higgs
physics \cite{Kribs, Hashimoto, Alwall}, electroweak precision tests \cite%
{Kribs, Alwall, Chanowitz, Novikov, Erler,Polonsky}, renormalisation group effects
\cite{Hung, Xiong} and flavor physics \cite{Babu, London, Dincer, Arhrib,
Hou1, Hou2, Alieve, Bashiry, soniAlok, Herrera, Bobrowski, Eilam, Giri,
Buras1, Hou3, lunghinew}. In addition to this the detailed analysis of supersymmetry in
the presence of a fourth generation have recently been performed in \cite%
{susy1, susy2}.

In flavor Physics the importance of SM4 is in the Flavor Changing Neutral
Currents (FCNC) which lies in the fact that on one hand it contains much
fewer parameters than other New Physics (NP) scenarios like the Littlest
Higgs model with T-parity (LHT), Randall-Sundrum (RS) models or the general
Minimal Supersymmetric Standard Model (MSSM) and on the other hand there is
the possibility of having simultaneously sizable NP physics effects in the $K
$ and $B$ systems compared to above mentioned NP models.  Moreover, having
the same operator structure as of the SM3, it implies that the
nonperturbative uncertainties in the SM4 are at the same level as in the
SM3. Recently, Buras et al. \cite{Buras1} have performed a detailed analysis
of non-Minimal Flavor Violating (MFV) effects in the $K$, $B_{d}$ and $B_{s}$
system in the SM4 where they paid particular attention to the correlation
between flavor observables and addressed with in this framework a number of
anomalies present in the experimental data. In addition to this they have
also studied the $D^{0}-\bar{D}^{0}$ mixing in the SM4 where they calculated
the size of allowed CP violation which is found at the observable level well
beyond anything possible with CKM dynamics \cite{Duling}.

In this work we investigate the possibility of searching for NP in the $\bar{%
B}_{0}\rightarrow K_{0}^{\ast }(1430)l^{+}l^{-}$( $l$ = $\mu $, $\tau $),
where $K_{0}^{\ast }(1430)$ is a scalar meson, using the fourth generation
of quarks $\left( t^{\prime }\text{, }b^{\prime }\right) $. At quark level
this decay is governed by $b\rightarrow s$ transitions which are in
forefront of indirect investigation of fourth generation. In these FCNC
transitions the fourth generation quark $\left( t^{\prime }\right) $, like $u
$, $c$, $t$ quarks, contributes at loop level. Therefore, it modifies the
corresponding Wilson coefficients which may have effects on branching ratio
and lepton polarization asymmetries of $\bar{B}_{0}\rightarrow K_{0}^{\ast
}(1430)l^{+}l^{-}$ decay. Now the main job of investigating the
semi-leptonic $B$ meson decay is to properly evaluate the hadronic matrix
elements for $B\rightarrow K_{0}^{\ast }(1430)$, namely the transition form
factors, which are governed by the non-perturbative QCD dynamics. Several
methods exist in the literature to deal with this problem, such as simple
quark model, light-front approach, QCD Sum Rules (QCDSR), light-cone QCD sum
rules (LCSR) and perturbative QCD factorization approach (PQCD).

In our numerical analysis for $\bar{B}_{0}\rightarrow K_{0}^{\ast }(1430)$
decays, we shall use the results of the form factors calculated by LCSR
approach in Ref. \cite{YuMing}, and explore the effects of fourth generation
parameters $\left( m_{t^{\prime }}\text{, }V_{t^{\prime }b}^{\ast
}V_{t^{\prime }s}\right) $ on branching ratios and lepton polarization
asymmetries. By incorporating the recent constraints $m_{t^{\prime
}}=400-600 $ GeV and $V_{t^{\prime }b}V_{t^{\prime }s}=\left(
0.05-1.4\right) \times 10^{-2}$ \cite{Giri} our results show that the decay
rates are quite sensitive to these parameters. Now the forward-backward
asymmetry is zero in the SM3 for these decays because of the absence of the
scalar type coupling, and it remains zero in SM4 as there is no new operator
in addition to the SM3 operators. The hadronic uncertainties associated with
the form factors and other input parameters have negligible effects on the
lepton polarization asymmetries and this makes them the efficient way in
establishing the NP. Here, we have also studied these asymmetries in the SM4
found that the effects of fourth generation parameters are quite significant
in some regions of parameter space of SM4.

The paper is organized as follows. In Sec. II, we present the effective
Hamiltonian for the semileptonic decay $\bar{B}\rightarrow K_{0}^{\ast
}l^{+}l^{-}$
Section III contains the parameterizations and numbers of the form factors
for the said decay using the LCSR approach. In Sec. IV we present the basic
formulas of physical observables like decay rates and polarization
asymmetries of final state lepton for the said decay. Section V is devoted
to the numerical analysis where we study the sensitivity of these physical
observables on fourth generation parameter $\left( m_{t^{\prime }}\text{, }%
V_{t^{\prime }b}^{\ast }V_{t^{\prime }s}\right) $. The main results are
summarized in Sec. VI.

\section{Effective Hamiltonia}

At quark level the decay $B\rightarrow K_{0}^{\ast }(1430)l^{+}l^{-}$ is
governed by the transition $b\rightarrow \ sl^{+}l^{-}$ for which the
effective Hamiltonian can be written as
\begin{equation}
H_{eff}=-\frac{4G_{F}}{\sqrt{2}}V_{tb}^{\ast }V_{ts}{\sum\limits_{i=1}^{10}}%
C_{i}({\mu })O_{i}({\mu }),  \label{effective hamiltonian 1}
\end{equation}%
where $O_{i}({\mu })$ $(i=1,\ldots ,10)$ are the four-quark operators and $%
C_{i}({\mu })$ are the corresponding Wilson\ coefficients at the energy
scale ${\mu }$ and the explicit expressions of these in the SM3 at NLO and
NNLL are given in \cite{Buchalla, Buras, Kim, Ali, Kruger,Grinstein, Cella,
Bobeth, Asatrian, Misiak, Huber}. The operators responsible for $%
B\rightarrow K_{0}^{\ast }(1430)l^{+}l^{-}$ are $O_{7}$, $O_{9}$ and $O_{10}$
and their form is given by
\begin{eqnarray}
O_{7} &=&\frac{e^{2}}{16\pi ^{2}}m_{b}\left( \bar{s}\sigma _{\mu \nu
}P_{R}b\right) F^{\mu \nu },\,  \notag \\
O_{9} &=&\frac{e^{2}}{16\pi ^{2}}(\bar{s}\gamma _{\mu }P_{L}b)(\bar{l}\gamma
^{\mu }l),\,  \label{op-form} \\
O_{10} &=&\frac{e^{2}}{16\pi ^{2}}(\bar{s}\gamma _{\mu }P_{L}b)(\bar{l}%
\gamma ^{\mu }\gamma _{5}l),  \notag
\end{eqnarray}%
with $P_{L,R}=\left( 1\pm \gamma _{5}\right) /2$. Now, the fourth generation
comes into the play just in the same way as the three generation SM, i.e.
the full set of operators remains the same as in the SM3. Therefore, the
effect of fourth generation displays itself by changing the values of Wilson
coefficients $C_{7}\left( \mu \right) $, $C_{9}\left( \mu \right) $ and $%
C_{10}$ via the virtual exchange of fourth generation up-type quark $%
t^{\prime }$ which then takes the form;%
\begin{equation}
\lambda _{t}C_{i}\rightarrow \lambda _{t}C_{i}^{SM}+\lambda _{t^{\prime
}}C_{i}^{new},  \label{wilson-modified}
\end{equation}%
where $\lambda _{f}=V_{fb}^{\ast }V_{fs}$ and the explicit forms of the $%
C_{i}$'s can be obtained from the corresponding expressions of the Wilson
coefficients in SM3 by substituting $m_{t}\rightarrow m_{t^{\prime }}$. By
adding an extra family of quarks, the CKM matrix of SM3 is extended by
another row and column which now becomes $4\times 4$. The unitarity of which
leads to%
\begin{equation*}
\lambda _{u}+\lambda _{c}+\lambda _{t}+\lambda _{t^{\prime }}=0.
\end{equation*}%
Since $\lambda _{u}=V_{ub}^{\ast }V_{us}$ has a very small value compared to
the others, therefore, we will ignore it. Then $\lambda _{t}\approx -\lambda
_{c}-\lambda _{t^{\prime }}$ and from Eq. (\ref{wilson-modified}) we have%
\begin{equation}
\lambda _{t}C_{i}^{SM}+\lambda _{t^{\prime }}C_{i}^{new}=-\lambda
_{c}C_{i}^{SM}+\lambda _{t^{\prime }}\left( C_{i}^{new}-C_{i}^{SM}\right) .
\label{wilson-modified1}
\end{equation}%
One can clearly see that under $\lambda _{t^{\prime }}\rightarrow 0$ or $%
m_{t^{\prime }}\rightarrow m_{t}$ the term $\lambda _{t^{\prime }}\left(
C_{i}^{new}-C_{i}^{SM}\right) $ vanishes which is the requirement of GIM
mechanism. Taking the contribution of the $t^{\prime }$ quark in the loop
the Wilson coefficients $C_{i}$'s can be written in the following form%
\begin{eqnarray}
C_{7}^{tot}\left( \mu \right)  &=&C_{7}^{SM}\left( \mu \right) +\frac{%
\lambda _{t^{\prime }}}{\lambda _{t}}C_{7}^{new}\left( \mu \right) ,  \notag
\\
C_{9}^{tot}\left( \mu \right)  &=&C_{9}^{SM}\left( \mu \right) +\frac{%
\lambda _{t^{\prime }}}{\lambda _{t}}C_{9}^{new}\left( \mu \right) ,
\label{wilson-tot} \\
C_{10}^{tot} &=&C_{10}^{SM}+\frac{\lambda _{t^{\prime }}}{\lambda _{t}}%
C_{10}^{new},  \notag
\end{eqnarray}%
where we factored out $\lambda _{t}=V_{tb}^{\ast }V_{ts}$ term in the
effective Hamiltonian given in Eq. (\ref{effective hamiltonian 1}) and the
last term in these expressions corresponds to the contribution of the $%
t^{\prime }$ quark to the Wilson Coefficients. $\lambda _{t^{\prime }}$ can
be parameterized as:
\begin{equation}
\lambda _{t^{\prime }}=\left\vert V_{t^{\prime }b}^{\ast }V_{t^{\prime
}s}\right\vert e^{i\phi _{sb}}
\end{equation}%
. In terms of the above Hamiltonian, the free quark decay amplitude for $%
b\rightarrow s$ $l^{+}l^{-}$ in SM4 can be derived as:
\begin{eqnarray}
\mathcal{M}(b &\rightarrow &sl^{+}l^{-})=-\frac{G_{F}\alpha }{\sqrt{2}\pi }%
V_{tb}V_{ts}^{\ast }\bigg\{C_{9}^{tot}(\bar{s}\gamma _{\mu }P_{L}b)(\bar{l}%
\gamma ^{\mu }l)+C_{10}^{tot}(\bar{s}\gamma _{\mu }P_{L}b)(\bar{l}\gamma
^{\mu }\gamma _{5}l)  \notag \\
&&-2m_{b}C_{7}^{tot}(\bar{s}i\sigma _{\mu \nu }\frac{q^{\nu }}{q^{2}}P_{R}b)(%
\bar{l}\gamma ^{\mu }l)\bigg\},  \label{quark-amplitude}
\end{eqnarray}%
where $q^{2}$ is the square of momentum transfer. The operator $O_{10}$ can
not be induced by the insertion of four-quark operators because of the
absence of the $Z$ -boson in the effective theory. Therefore, the Wilson
coefficient $C_{10}$ does not renormalize under QCD corrections and hence it
is independent on the energy scale. In addition to this, the above quark
level decay amplitude can receive contributions from the matrix element of
four-quark operators, $\sum_{i=1}^{6}\langle l^{+}l^{-}s|O_{i}|b\rangle $,
which are usually absorbed into the effective Wilson coefficient $%
C_{9}^{SM}(\mu )$ and can usually be called $C_{9}^{eff}$, that one can
decompose into the following three parts
\begin{equation*}
C_{9}^{SM}=C_{9}^{eff}(\mu )=C_{9}(\mu )+Y_{SD}(z,s^{\prime
})+Y_{LD}(z,s^{\prime }),
\end{equation*}%
where the parameters $z$ and $s^{\prime }$ are defined as $%
z=m_{c}/m_{b},\,\,\,s^{\prime }=q^{2}/m_{b}^{2}$. $Y_{SD}(z,s^{\prime })$
describes the short-distance contributions from four-quark operators far
away from the $c\bar{c}$ resonance regions, which can be calculated reliably
in the perturbative theory. The long-distance contributions $%
Y_{LD}(z,s^{\prime })$ from four-quark operators near the $c\bar{c}$
resonance cannot be calculated from first principles of QCD and are usually
parameterized in the form of a phenomenological Breit-Wigner formula making
use of the vacuum saturation approximation and quark-hadron duality. We will
neglect the long-distance contributions in this work because of the absence
of experimental data on $B\rightarrow J/\psi K_{0}^{\ast }(1430)$. The
manifest expressions for $Y_{SD}(z,s^{\prime })$ can be written as \cite%
{Buras}
\begin{eqnarray}
Y_{SD}(z,s^{\prime }) &=&h(z,s^{\prime })(3C_{1}(\mu )+C_{2}(\mu
)+3C_{3}(\mu )+C_{4}(\mu )+3C_{5}(\mu )+C_{6}(\mu ))  \notag \\
&&-\frac{1}{2}h(1,s^{\prime })(4C_{3}(\mu )+4C_{4}(\mu )+3C_{5}(\mu
)+C_{6}(\mu ))  \notag \\
&&-\frac{1}{2}h(0,s^{\prime })(C_{3}(\mu )+3C_{4}(\mu ))+{\frac{2}{9}}%
(3C_{3}(\mu )+C_{4}(\mu )+3C_{5}(\mu )+C_{6}(\mu )),  \label{short-distance}
\end{eqnarray}%
with
\begin{eqnarray}
h(z,s^{\prime }) &=&-{\frac{8}{9}}\mathrm{ln}z+{\frac{8}{27}}+{\frac{4}{9}}x-%
{\frac{2}{9}}(2+x)|1-x|^{1/2}\left\{
\begin{array}{l}
\ln \left\vert \frac{\sqrt{1-x}+1}{\sqrt{1-x}-1}\right\vert -i\pi \quad
\mathrm{for}{{\ }x\equiv 4z^{2}/s^{\prime }<1} \\
2\arctan \frac{1}{\sqrt{x-1}}\qquad \mathrm{for}{{\ }x\equiv
4z^{2}/s^{\prime }>1}%
\end{array}%
\right. ,  \notag \\
h(0,s^{\prime }) &=&{\frac{8}{27}}-{\frac{8}{9}}\mathrm{ln}{\frac{m_{b}}{\mu
}}-{\frac{4}{9}}\mathrm{ln}s^{\prime }+{\frac{4}{9}}i\pi \,\,.  \label{hzs}
\end{eqnarray}

Apart from this, the non-factorizable effects \cite{b to s 1, b to s 2, b to
s 3,NF charm loop} from the charm loop can bring about further corrections
to the radiative $b\rightarrow s\gamma $ transition, which can be absorbed
into the effective Wilson coefficient $C_{7}^{eff}$. Specifically, the
Wilson coefficient $C^{eff}_{7}$ is given by \cite{c.q. geng 4}
\[
C_{7}^{SM}(\mu )=C_{7}^{eff}(\mu )=C_{7}(\mu )+C_{b\rightarrow s\gamma }(\mu ),
\]
with
\begin{eqnarray}
C_{b\rightarrow s\gamma }(\mu ) &=&i\alpha _{s}\bigg[{\frac{2}{9}}\eta
^{14/23}(G_{1}(x_{t})-0.1687)-0.03C_{2}(\mu )\bigg], \\
G_{1}(x_{t}) &=&{\frac{x_{t}(x_{t}^{2}-5x_{t}-2)}{8(x_{t}-1)^{3}}}+{\frac{3x_{t}^{2}\mathrm{ln}^{2}x_{t}%
}{4(x_{t}-1)^{4}}},
\end{eqnarray}
where $\eta =\alpha _{s}(m_{W})/\alpha _{s}(\mu )$, $%
x_{t}=m_{t}^{2}/m_{W}^{2}$, $C_{b\rightarrow s\gamma }$ is the absorptive
part for the $b\rightarrow sc\bar{c}\rightarrow s\gamma $ rescattering and
we have dropped out the tiny contributions proportional to CKM sector $%
V_{ub}V_{us}^{\ast }$. In addition, $C_{7}^{new}(\mu )$ can be obtained by replacing
$m_{t}$ with $m_{t\prime}$ in the above expression. Similar replacement $(m_{t} \to m_{t\prime})$
has to be done for the other Wilson Coefficients
$C^{eff}_9$ and $C_{10}$ which have too lengthy expressions to give here and their explicit expressions
are given in refs. \cite{Buchalla, Buras, Kim, Ali, Kruger,Grinstein, Cella,
Bobeth, Asatrian, Misiak, Huber}.

\section{Parameterizations of matrix elements and form factors in LCSR}

With the free quark decay amplitude available, we can proceed to calculate
the decay amplitudes for semi-leptonic decays of $\bar{B}_{0}\rightarrow
K_{0}^{\ast }(1430)l^{+}l^{-}$ at hadronic level, which can be obtained by
sandwiching the free quark amplitudes between the initial and final meson
states. Consequently, the following two hadronic matrix elements
\begin{equation*}
\langle K_{0}^{\ast }(p)|\bar{s}\gamma _{\mu }\gamma _{5}b|B_{q^{\prime
}}(p+q)\rangle ,\,\,\,\langle K_{0}^{\ast }(p)|\bar{s}\sigma _{\mu \nu
}\gamma _{5}q^{\nu }b|B_{q^{\prime }}(p+q)\rangle
\end{equation*}%
need to be computed as can be observed from Eq. (\ref{effective hamiltonian
1}). Generally, the above two matrix elements can be parameterized in terms
of a series of form factors as
\begin{eqnarray}
\langle K_{0}^{\ast }(p)|\bar{s}\gamma _{\mu }\gamma _{5}b|B_{q^{\prime
}}(p+q)\rangle  &=&-i[f_{+}(q^{2})p_{\mu }+f_{-}(q^{2})q_{\mu }],
\label{axial form factor} \\
\langle K_{0}^{\ast }(p)|\bar{s}\sigma _{\mu \nu }\gamma _{5}q^{\nu
}b|B_{q^{\prime }}(p+q)\rangle  &=&-\frac{1}{m_{B}+m_{K_{0}^{\ast }}}\left[
\left( 2p+q\right) _{\mu }q^{2}-\left( m_{B}^{2}-m_{K_{0}^{\ast
}}^{2}\right) q_{\mu }\right] f_{T}\left( q^{2}\right) .
\label{tensor form factor}
\end{eqnarray}

The form factors are the non-perturbative quantities and to calculate them
one has to rely on some non-perturbative approaches. Considering the
distribution amplitudes up to twist-3, the form factors at small $q^{2}$ for
$\bar{B}_{0}\rightarrow K_{0}^{\ast }l^{+}l^{-}$ have been calculated in
\cite{YuMing} using the LCSR. The dependence of form factors $%
f_{i}(q^{2})(i=+,-,T)$ on momentum transfer $s$ are parameterized in either
the single pole form
\begin{equation}
f_{i}(q^{2})={\frac{f_{i}(0)}{1-a_{i}q^{2}/m_{B_{0}}^{2}}},  \label{form}
\end{equation}%
or the double-pole form
\begin{equation}
f_{i}(q^{2})={\frac{f_{i}(0)}{%
1-a_{i}q^{2}/m_{B_{0}}^{2}+b_{i}q^{4}/m_{B_{0}}^{4}}},  \label{form2}
\end{equation}%
in the whole kinematical region $0<q^{2}<(m_{B_{0}}-m_{K_{0}^{\ast }})^{2}$
while non-perturbative parameters $a_{i}$ and $b_{i}$ can be fixed by the
magnitudes of form factors corresponding to the small momentum transfer
calculated in the LCSR approach. The results for the parameters $a_{i}$, $%
b_{i}$ accounting for the $q^{2}$ dependence of form factors $f_{+}$, $f_{-}$
and $f_{T}$ are grouped in Table \ref{di-fit B to Kstar0(1430)}.

\begin{table}[htb]
\caption{Numerical results for the parameters $f_i(0)$, $a_i$ and $b_i$
involved in the double-pole fit of form factors (\protect\ref{form2})
responsible for $\bar{B}_0 \to K^{\ast}_0(1430) l^{+} l^{-}$ decay up to the
twist-3 distribution amplitudes of $K^{\ast}_0(1430) $ meson.}
\label{di-fit B to Kstar0(1430)}%
\begin{tabular}{cccc}
\hline\hline
& $\hspace{2 cm} f_i(0)$ & $\hspace{2 cm} a_i$ & $\hspace{2 cm} b_i$ \\
\hline
$f_{+}$ & $\hspace{2 cm} 0.97^{+0.20}_{-0.20}$ & \hspace{2 cm} $%
0.86^{+0.19}_{-0.18}$ &  \\ \hline
$f_{-}$ & $\hspace{2 cm} 0.073^{+0.02}_{-0.02}$ & \hspace{2 cm} $%
2.50^{+0.44}_{-0.47}$ & \hspace{2 cm} $1.82^{+0.69}_{-0.76}$ \\ \hline
$f_{T}$ & $\hspace{2 cm} 0.60^{+0.14}_{-0.13}$ & \hspace{2 cm} $%
0.69^{+0.26}_{-0.27}$ &  \\ \hline\hline
&  &  &
\end{tabular}%
\end{table}

\section{Formula for Physical Observables}

In this section, we are going to perform the calculations of some
interesting observables in phenomenology like the decay rates,
forward-backward asymmetry as well as the polarization asymmetries of final
state lepton. From Eq. (\ref{quark-amplitude}), it is straightforward to
obtain the decay amplitude for $\bar{B}_{0}\rightarrow K_{0}^{\ast
}l^{+}l^{-}$ as
\begin{equation}
\mathcal{M}_{\bar{B}_{0}\rightarrow K_{0}^{\ast }l^{+}l^{-}}=-\frac{%
G_{F}\alpha }{2\sqrt{2}\pi }V_{tb}V_{ts}^{\ast }\left[ T_{\mu }^{1}(\bar{l}%
\gamma ^{\mu }l)+T_{\mu }^{2}(\bar{l}\gamma ^{\mu }\gamma _{5}l)\right] ,
\label{lambda-amplitude}
\end{equation}%
where the functions $T_{\mu }^{1}$ and $T_{\mu }^{2}$ are given by
\begin{equation}
T_{\mu }^{1}=iC_{9}^{tot}f_{+}(q^{2})p_{\mu }+\frac{4im_{b}}{%
m_{B}+m_{K_{0}^{\ast }}}C_{7}^{tot}f_{T}(q^{2})p_{\mu },
\label{(first-aux-function)}
\end{equation}%
\begin{equation*}
T_{\mu }^{2}=iC_{10}^{tot}\left( f_{+}(q^{2})p_{\mu }+f_{-}(q^{2})q_{\mu
}\right) ,
\end{equation*}%
Due to the equation of motion for lepton fields, the terms proportional to $%
q_{\mu }$ in $T_{\mu }^{1}$, namely $f_{-}(q^{2})$ do not contribute to the
decay amplitude.

\subsection{The differential decay rates and forward-backward asymmetry of $%
\bar{B}_0\rightarrow K^{\ast}_0(1430) l^{+}l^{-}$}

The semi-leptonic decay $\bar{B}_{0}\rightarrow K_{0}^{\ast }(1430)l^{+}l^{-}
$ is induced by FCNCs. The differential decay width of $\bar{B}%
_{0}\rightarrow K_{0}^{\ast }(1430)l^{+}l^{-}$ in the rest frame of $\bar{B}%
_{0}$ meson can be written as \cite{PDG}
\begin{equation}
{\frac{d\Gamma (\bar{B}_{0}\rightarrow K_{0}^{\ast }(1430)l^{+}l^{-})}{dq^{2}%
}}={\frac{1}{(2\pi )^{3}}}{\frac{1}{32m_{\bar{B}_{0}}}}%
\int_{u_{min}}^{u_{max}}|{\widetilde{M}}_{\bar{B}_{0}\rightarrow K_{0}^{\ast
}(1430)l^{+}l^{-}}|^{2}du,  \label{differential decay width}
\end{equation}%
where $u=(p_{K_{0}^{\ast }(1430)}+p_{l^{-}})^{2}$ and $%
q^{2}=(p_{l^{+}}+p_{l^{-}})^{2}$; $p_{K_{0}^{\ast }(1430)}$, $p_{l^{+}}$ and
$p_{l^{-}}$ are the four-momenta vectors of $K_{0}^{\ast }(1430)$, $l^{+}$
and $l^{-}$ respectively; $|{\widetilde{M}}_{\bar{B}_{0}\rightarrow
K_{0}^{\ast }(1430)l^{+}l^{-}}|^{2}$ is the squared decay amplitude after
integrating over the angle between the lepton $l^{-}$ and $K_{0}^{\ast
}(1430)$ meson. The upper and lower limits of $u$ are given by
\begin{eqnarray}
u_{max} &=&(E_{K_{0}^{\ast }(1430)}^{\ast }+E_{l^{-}}^{\ast })^{2}-(\sqrt{%
E_{K_{0}^{\ast }(1430)}^{\ast 2}-m_{K_{0}^{\ast }(1430)}^{2}}-\sqrt{%
E_{l^{-}}^{\ast 2}-m_{l^{-}}^{2}})^{2},  \notag \\
u_{min} &=&(E_{K_{0}^{\ast }(1430)}^{\ast }+E_{l^{-}}^{\ast })^{2}-(\sqrt{%
E_{K_{0}^{\ast }(1430)}^{\ast 2}-m_{K_{0}^{\ast }(1430)}^{2}}+\sqrt{%
E_{l^{-}}^{\ast 2}-m_{l^{-}}^{2}})^{2};
\end{eqnarray}%
where the energies of $K_{0}^{\ast }(1430)$ and $l^{-}$ in the rest frame of
lepton pair $E_{K_{0}^{\ast }(1430)}^{\ast }$ and $E_{l^{-}}^{\ast }$ are
determined as
\begin{equation}
E_{K_{0}^{\ast }(1430)}^{\ast }={\frac{m_{\bar{B}_{0}}^{2}-m_{K_{0}^{\ast
}(1430)}^{2}-q^{2}}{2\sqrt{q^{2}}}},\hspace{1cm}E_{l}^{\ast }={\frac{q^{2}}{2%
\sqrt{q^{2}}}}.
\end{equation}%
Collecting everything together, one can write the general expression of the
differential decay rate for $\bar{B}_{0}\rightarrow K_{0}^{\ast
}(1430)l^{+}l^{-}$ as:
\begin{eqnarray}
\frac{d\Gamma }{dq^{2}} &=&\frac{G_{F}^{2}\alpha ^{2}\left\vert
V_{tb}V_{ts}^{\ast }\right\vert ^{2}}{3072m_{B}^{3}\pi ^{5}q^{2}}\sqrt{1-%
\frac{4m_{l}^{2}}{q^{2}}}\sqrt{\lambda (m_{B}^{2},m_{K_{0}^{\ast
}}^{2},q^{2})}\times   \notag \\
&&\bigg\{\left\vert A\right\vert ^{2}\left( 2m_{l}^{2}+q^{2}\right) \lambda
+12q^{2}m_{l}^{2}\left( m_{B}^{2}-m_{K_{0}^{\ast }}^{2}-q^{2}\right) \left(
CB^{\ast }+C^{\ast }B\right) +12m_{l}^{2}q^{4}\left\vert C\right\vert ^{2}
\notag \\
&&+\left\vert B\right\vert ^{2}\left( \left( 2m_{l}^{2}+q^{2}\right) \left(
m_{B}^{4}-2m_{B}^{2}m_{K_{0}^{\ast }}^{2}-2q^{2}m_{K_{0}^{\ast }}^{2}\right)
+\left( m_{K_{0}^{\ast }}^{2}-q^{2}\right) ^{2}+2m_{l}^{2}\left(
m_{K_{0}^{\ast }}^{4}+10tm_{K_{0}^{\ast }}^{2}+q^{4}\right) \right) \bigg\},
\label{drate}
\end{eqnarray}%
where
\begin{equation}
\lambda =\lambda (m_{B}^{2},m_{K_{0}^{\ast
}}^{2},q^2)=m_{B}^{4}+m_{K_{0}^{\ast }}^{4}+q^{4}-2m_{B}^{2}m_{K_{0}^{\ast
}}^{2}-2m_{K_{0}^{\ast }}^{2}q^{2}-2q^{2}m_{B}^{2}.  \label{function1}
\end{equation}%
The auxiliary functions are defined as
\begin{eqnarray}
A &=&iC_{9}^{tot}f_{+}(q^{2})+\frac{4im_{b}}{m_{B}+m_{K_{0}^{\ast }}}%
C_{7}^{tot}f_{T}(q^{2})  \notag \\
B &=&iC_{10}^{tot}f_{+}(q^{2})  \notag \\
C &=&iC_{10}^{tot}f_{-}\left( q^{2}\right)   \label{nauxfunction}
\end{eqnarray}%
Just to make a comment, the form factor $f_{-}(q^{2})$ is an order of
magnitude smaller than the form factors $f_{+}(q^{2})$ and $f_{T}(q^{2})$,
therefore, the value of auxiliary function $C$ is suppressed by the same
magnitude compared to $A$ and $B$.

\subsection{Lepton Polarization asymmetries of $\bar{B}_0\rightarrow
K^{\ast}_0(1430) l^{+}l^{-}$}

In the rest frame of the lepton $l^{-}$, the unit vectors along
longitudinal, normal and transversal component of the $l^{-}$ can be defined
as \cite{Aliev UED}:
\begin{eqnarray}
s_{L}^{-\mu } &=&(0,\vec{e}_{L})=\left( 0,\frac{\vec{p}_{-}}{\left| \vec{p}%
_{-}\right| }\right) ,  \notag \\
s_{N}^{-\mu } &=&(0,\vec{e}_{N})=\left( 0,\frac{\vec{p}_{K^{\ast}_0 }\times
\vec{p}_{-}}{\left| \vec{p}_{K^{\ast}_0 }\times \vec{p}_{-}\right| }\right) ,
\label{p-vectors} \\
s_{T}^{-\mu } &=&(0,\vec{e}_{T})=\left( 0,\vec{e}_{N}\times \vec{e}%
_{L}\right) ,  \notag
\end{eqnarray}
where $\vec{p}_{-}$ and $\vec{p}_{K_{0}^{*}}$ are the three-momenta of the
lepton $l^{-}$ and $K_{0}^{*}(1430)$ meson respectively in the center mass
(CM) frame of $l^{+}l^{-}$ system. Lorentz transformation is used to boost
the longitudinal component of the lepton polarization to the CM frame of the
lepton pair as
\begin{equation}
\left( s_{L}^{-\mu }\right) _{CM}=\left( \frac{|\vec{p}_{-}|}{m_{l}},\frac{%
E_{l}\vec{p}_{-}}{m_{l}\left| \vec{p}_{-}\right| }\right)
\label{bossted component}
\end{equation}
where $E_{l}$ and $m_{l}$ are the energy and mass of the lepton. The normal
and transverse components remain unchanged under the Lorentz boost.

The longitudinal ($P_{L}$), normal ($P_{N}$) and transverse ($P_{T}$)
polarizations of lepton can be defined as:
\begin{equation}
P_{i}^{(\mp )}(q^{2})=\frac{\frac{d\Gamma }{dq^{2}}(\vec{\xi}^{\mp }=\vec{e}%
^{\mp })-\frac{d\Gamma }{dq^{2}}(\vec{\xi}^{\mp }=-\vec{e}^{\mp })}{\frac{%
d\Gamma }{dq^{2}}(\vec{\xi}^{\mp }=\vec{e}^{\mp })+\frac{d\Gamma }{dq^{2}}(%
\vec{\xi}^{\mp }=-\vec{e}^{\mp })}  \label{polarization-defination}
\end{equation}%
where $i=L,\;N,\;T$ and $\vec{\xi}^{\mp }$ is the spin direction along the
leptons $l^{\mp }$. The differential decay rate for polarized lepton $l^{\mp
}$ in $\bar{B}_{0}\rightarrow K_{0}^{\ast }(1430)l^{+}l^{-}$ decay along any
spin direction $\vec{\xi}^{\mp }$ is related to the unpolarized decay rate (%
\ref{differential decay width}) with the following relation
\begin{equation}
\frac{d\Gamma (\vec{\xi}^{\mp })}{dq^{2}}=\frac{1}{2}\left( \frac{d\Gamma }{%
dq^{2}}\right) [1+(P_{L}^{\mp }\vec{e}_{L}^{\mp }+P_{N}^{\mp }\vec{e}%
_{N}^{\mp }+P_{T}^{\mp }\vec{e}_{T}^{\mp })\cdot \vec{\xi}^{\mp }].
\label{polarized-decay}
\end{equation}%
We can achieve the expressions of longitudinal, normal and transverse
polarizations for $\bar{B}_{0}\rightarrow K_{0}^{\ast }(1430)l^{+}l^{-}$
decays as collected below. The longitudinal lepton polarization can be
written as \cite{YuMing}
\begin{equation}
P_{L}(q^{2})=(1/{\frac{d\Gamma }{dq^{2}}})\frac{\alpha
^{2}G_{F}^{2}\left\vert V_{tb}^{\ast }V_{ts}\right\vert ^{2}\lambda
^{3/2}(m_{B}^{2},m_{K_{0}^{\ast }}^{2},q^{2})}{3072m_{B}^{3}\pi ^{5}}(1-%
\frac{4m_{l}^{2}}{q^{2}})(AB^{\ast }+A^{\ast }B).
\label{longitudinal-polrization}
\end{equation}%
Similarly, the normal lepton polarization is
\begin{equation}
P_{N}(q^{2})=(1/{\frac{d\Gamma }{dq^{2}}})\frac{\alpha
^{2}G_{F}^{2}\left\vert V_{tb}^{\ast }V_{ts}\right\vert ^{2}m_{l}}{%
4096m_{B}^{3}\pi ^{4}\sqrt{q^{2}}}\sqrt{1-\frac{4m_{l}^{2}}{s}}\bigg[%
(m_{B}^{2}-m_{K_{0}^{\ast }}^{2}+q^{2})(A^{\ast }B+AB^{\ast
})-2q^{2}(A^{\ast }C+AC^{\ast })\bigg],  \label{Normal-polarization}
\end{equation}%
and the transverse one is given by
\begin{eqnarray}
P_{T}(q^{2}) &=&(1/{\frac{d\Gamma }{dq^{2}}})\frac{-i\alpha
^{2}G_{F}^{2}\left\vert V_{tb}^{\ast }V_{ts}\right\vert ^{2}\lambda
^{1/2}(m_{B}^{2},m_{K_{0}^{\ast }}^{2},q^{2})}{2048m_{B}^{3}\pi ^{4}}m_{l}(1-%
\frac{4m_{l}^{2}}{q^{2}})(m_{B}^{2}-m_{K_{0}^{\ast }}^{2}+q^{2})(B^{\ast
}C-BC^{\ast }).  \notag \\
&&  \label{expression-TPolarization}
\end{eqnarray}%
The ${\frac{d\Gamma }{dq^{2}}}$ appearing in the above equation is the one
given in Eq. (\ref{drate}) and $\lambda (m_{B}^{2},m_{K_{0}^{\ast
}}^{2},q^{2})$ is the same as that defined in Eq. (\ref{function1}).

\section{Numerical Analysis}

In this section we will analyze the dependency of the total branching ratios
and different lepton polarizations on the fourth generation SM parameters
i.e. fourth generation quark mass ($m_{t^{\prime }}$) and to the product of
quark mixing matrix $V_{t^{\prime }b}^{\ast }V_{t^{\prime }s}=\left\vert
V_{t^{\prime }b}^{\ast }V_{t^{\prime }s}\right\vert e^{i\phi _{sb}}$. One of
the main input parameters is the form factors which are the non-perturbative
quantities and one needs some model to calculate them. Here, we will use the
form factors that were calculated using the LCSR \cite{YuMing} and their
dependence on $q^2$ is given in Section II and the corresponding values of
different parameters is listed in Table I. Also we use the next-to-leading
order approximation for the Wilson coefficients $C_{i}^{SM}$ and $C_{i}^{new}
$\cite{Buras, Asatrian} at the renormalization point $\mu =m_{b}$. It has
already been mentioned that besides \ the short distance contributions in
the $C_{9}^{eff}$ there are the long distance contributions resulting from
the $c\bar{c}$ resonances like $J/\Psi $ and its excited states. In the
present study we do not take these long distance effects into account and
also we use the cental value of the form factors and the other input parameters
given in Table I.
\begin{figure}[h]
\begin{center}
\begin{tabular}{cc}
\vspace{-2cm} \includegraphics[scale=0.6]{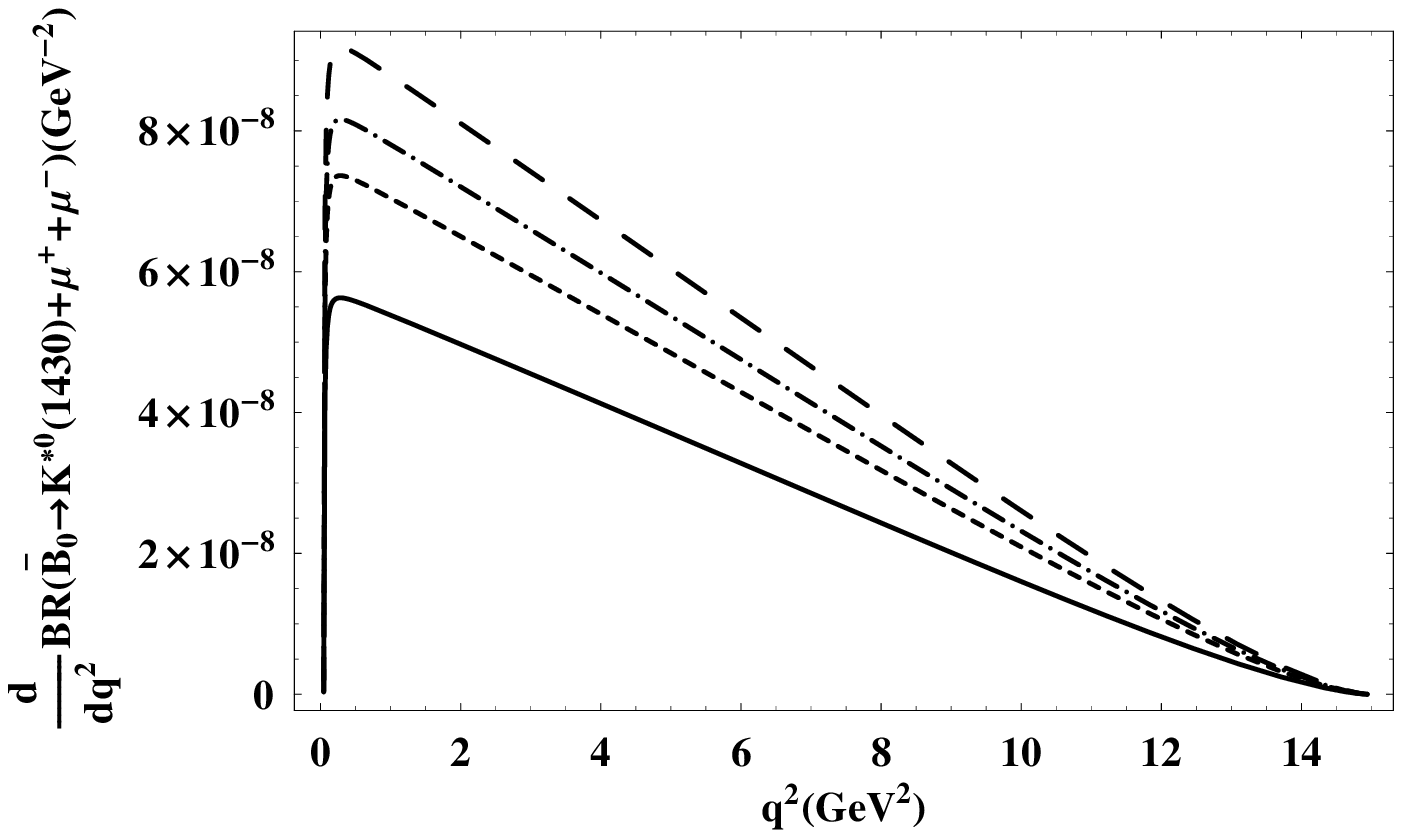} %
\includegraphics[scale=0.6]{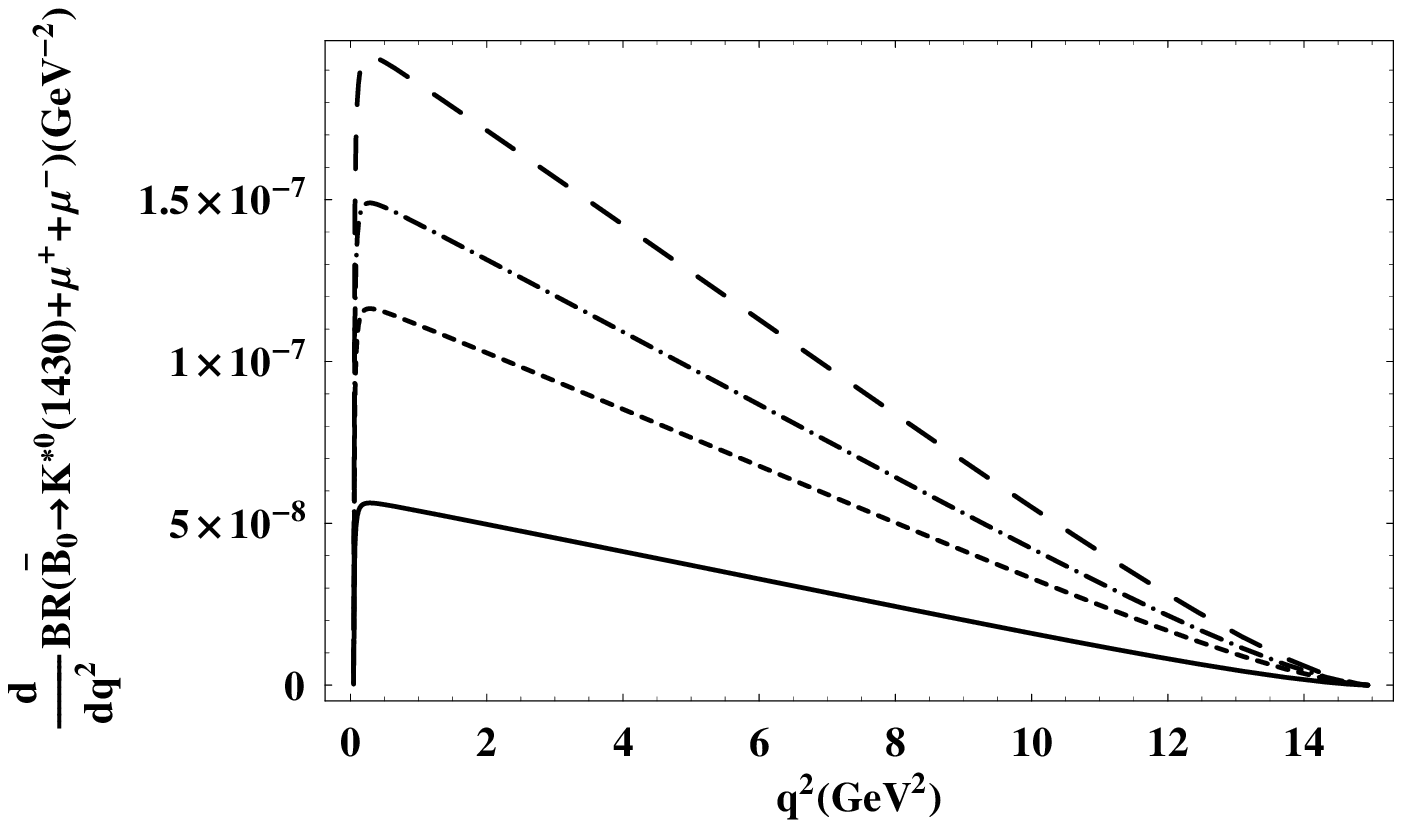} \put (-350,220){(a)} \put
(-100,220){(b)} &  \\
\includegraphics[scale=0.6]{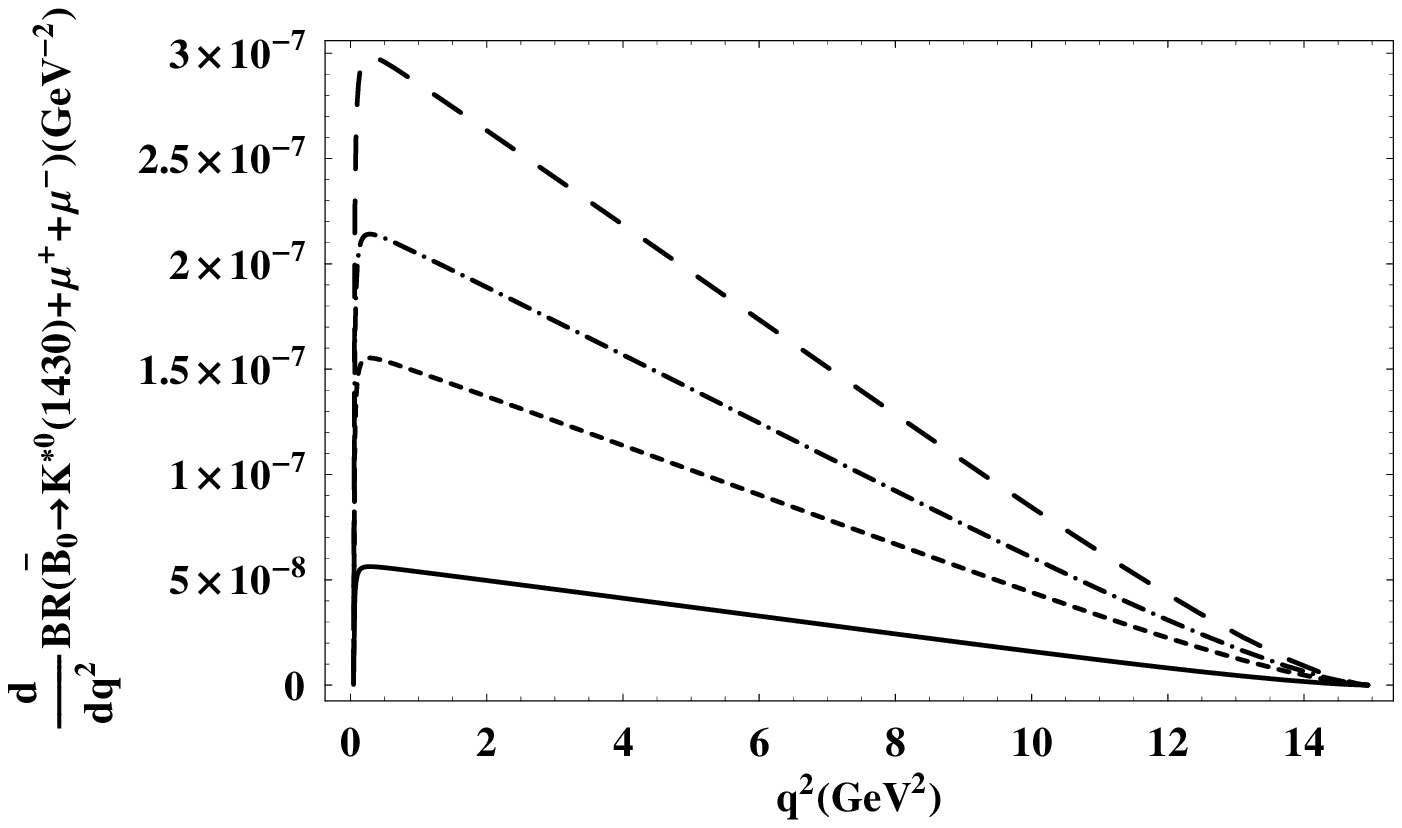} \includegraphics[scale=0.6]{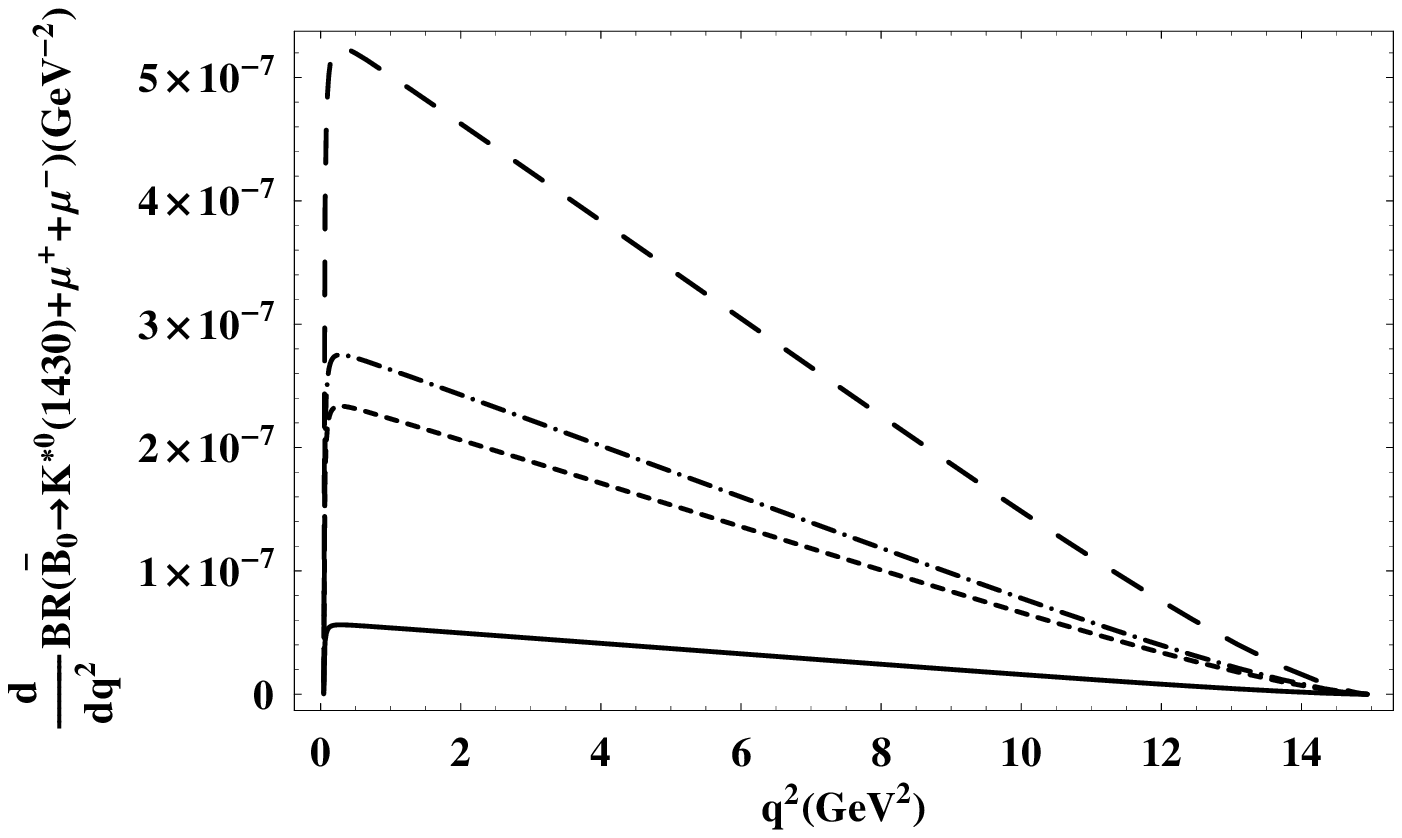}
\put (-350,220){(c)} \put (-100,220){(d)} &
\end{tabular}%
\end{center}
\caption{The dependence of branching ratio of $\bar{B}_{0}\rightarrow K_{0}^{\ast
}(1430)\protect\mu^{+}\protect\mu^{-}$ on $q^2$ for different values of $%
m_{t^{\prime }}$ and $\left\vert V_{t^{\prime }b}^{\ast }V_{t^{\prime
}s}\right\vert $. $\left\vert V_{t^{\prime }b}^{\ast }V_{t^{\prime
}s}\right\vert $ $=$ $0.002$, $0.006$, $0.009$ and $0.014$ in (a), (b), (c)
and (d) respectively. In all the graphs, the solid line corresponds to the
SM, dashed line, dashed-dotted and long dashed lines are for $m_{t^{\prime
}} $ $=$ $200$ GeV, $400$ GeV and $600$ GeV, respectively.}
\label{Branching ratio muons}
\end{figure}
\begin{figure}[h]
\begin{center}
\begin{tabular}{cc}
\vspace{-2cm} \includegraphics[scale=0.6]{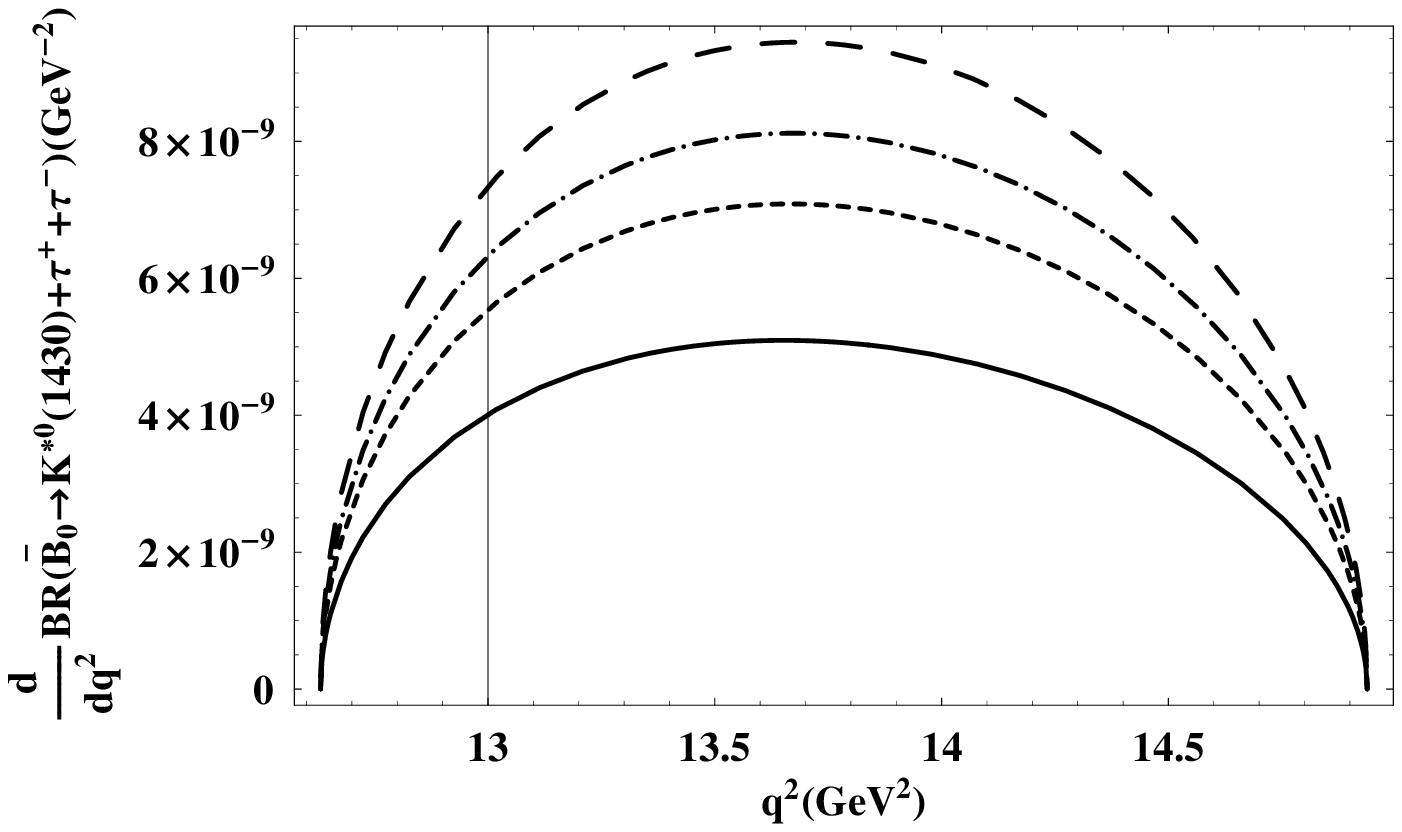} %
\includegraphics[scale=0.6]{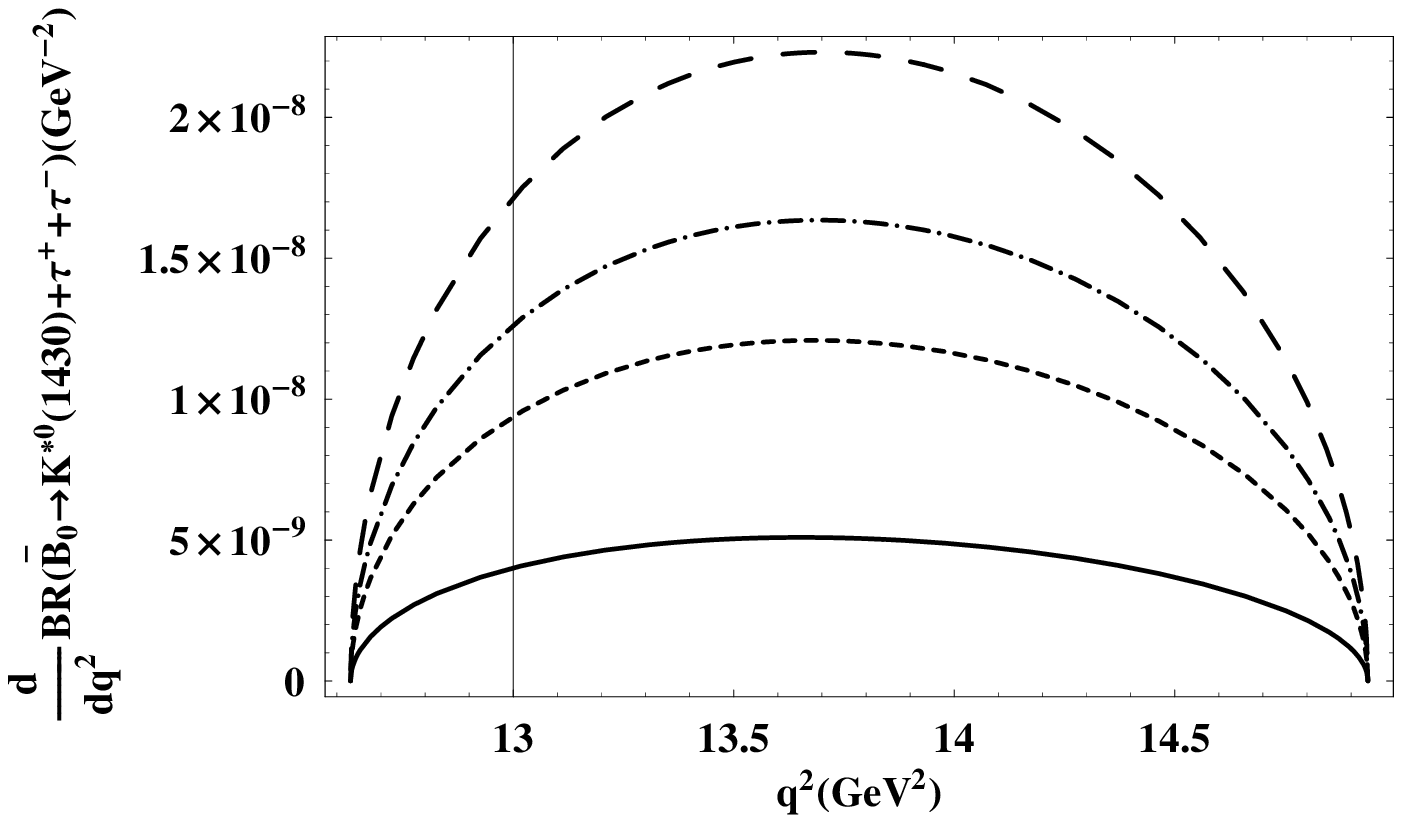} \put (-350,220){(a)} \put
(-100,220){(b)} &  \\
\includegraphics[scale=0.6]{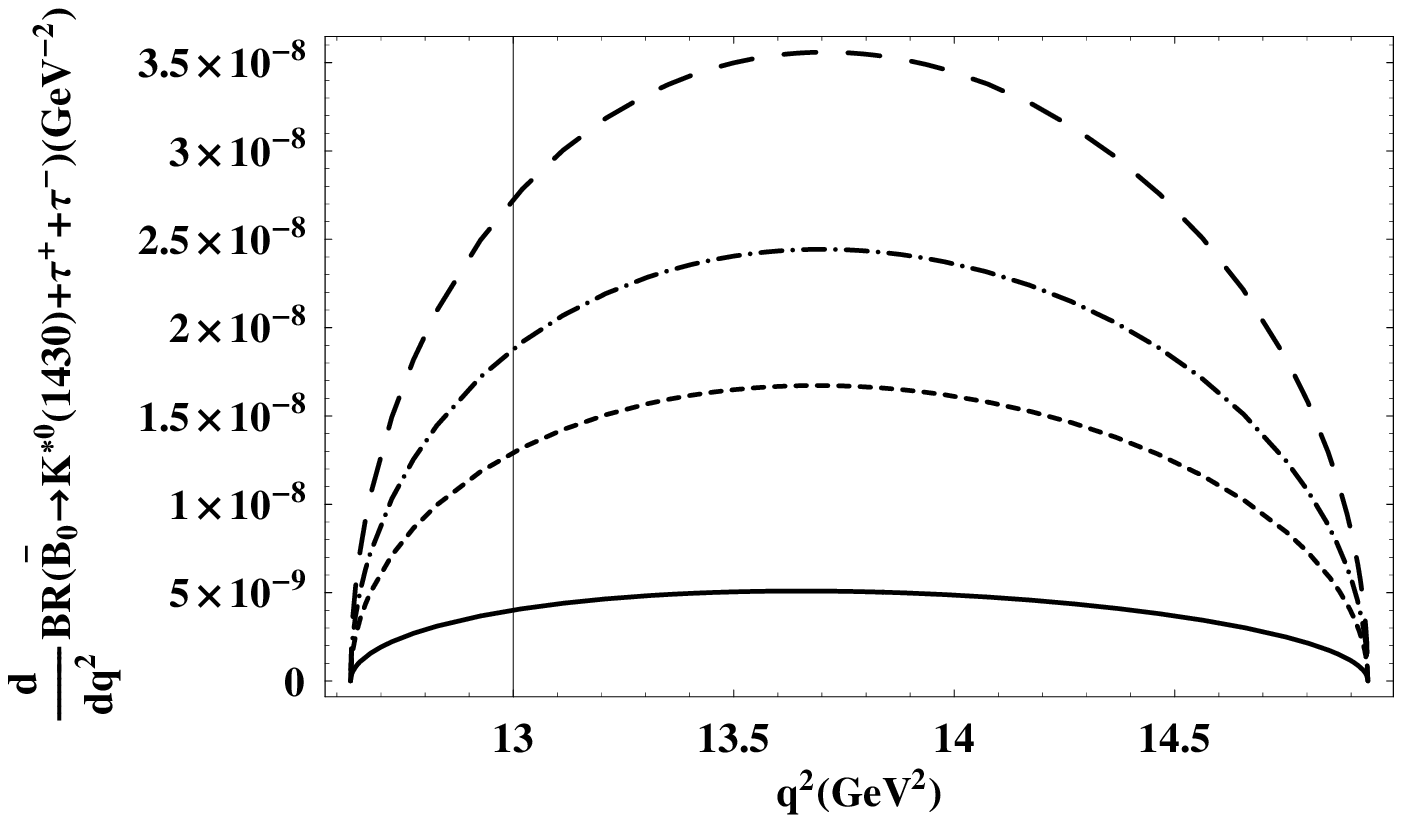} \includegraphics[scale=0.6]{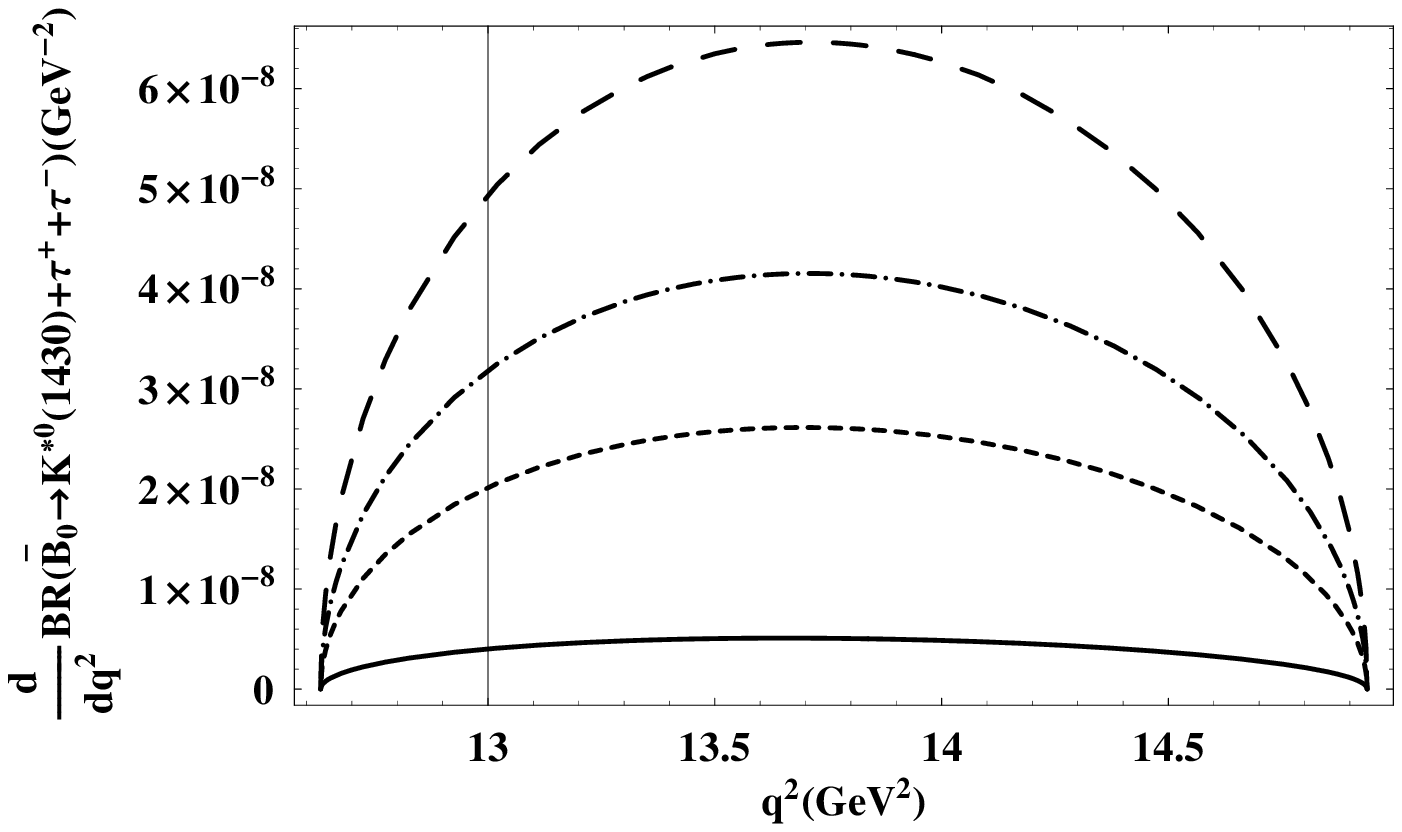}
\put (-350,220){(c)} \put (-100,220){(d)} &
\end{tabular}%
\end{center}
\caption{The dependence of branching ratio of $\bar{B}_{0}\rightarrow K_{0}^{\ast
}(1430)\protect\tau^{+}\protect\tau^{-}$ on $q^2$ for different values of $%
m_{t^{\prime }}$ and $\left\vert V_{t^{\prime }b}^{\ast }V_{t^{\prime
}s}\right\vert $. The values of fourth generation parameters and the legends
are same as in Fig.1}
\label{Branching ratio muons}
\end{figure}

In order to perform quantitative analysis of physical observables, it is
necessary to have the numerical values of the new parameters $\left(
m_{t^{\prime }}\text{, }\left\vert V_{t^{\prime }b}^{\ast }V_{t^{\prime
}s}\right\vert \text{, }\phi _{sb}\right) $. In the forthcoming analysis we
use the constraints of Ref. \cite{Giri} on the fourth generation
parameters, where it is found that $m_{t^{\prime }}$ varies from $400-600$
GeV with the mixing angle $\left\vert V_{t^{\prime }b}^{\ast }V_{t^{\prime
}s}\right\vert $ in the range of about $\left( 0.05\text{ to }1.4\right)
\times 10^{-2}$ and the value of CP-odd phase is from $0^{\circ }$ to $%
80^{\circ }$. Keeping the value of the phase $\phi _{sb}=$ $80^{\circ }$ and
for different values of $m_{t^{\prime }}$ and $\left\vert V_{t^{\prime
}b}^{\ast }V_{t^{\prime }s}\right\vert $ we will plot the physical
observables with square of the momentum transfer $q^{2}$ to see their
effects at small and large value of $q^{2}$.
\begin{figure}[h]
\begin{center}
\begin{tabular}{cc}
\vspace{-2cm} \includegraphics[scale=0.6]{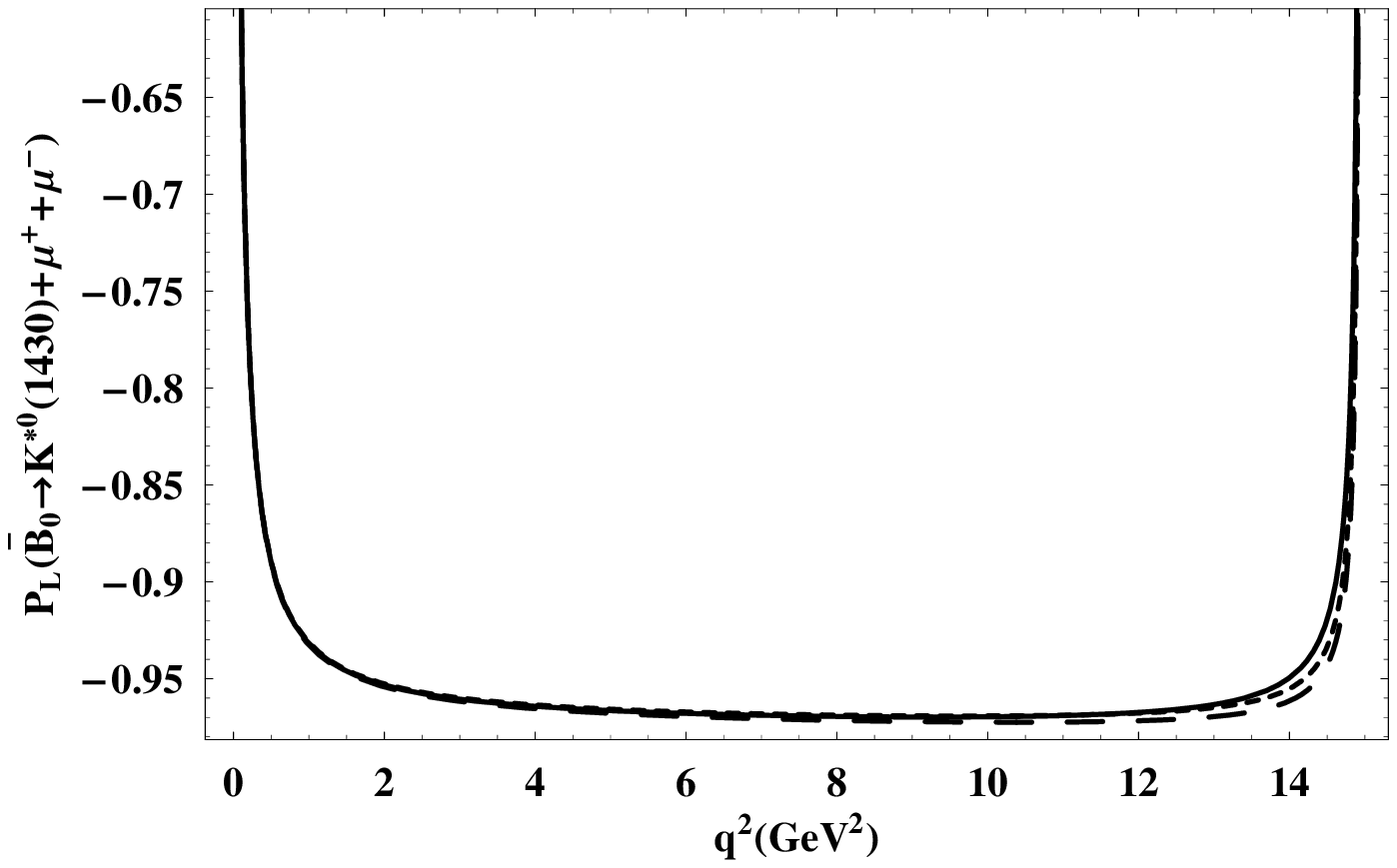} %
\includegraphics[scale=0.6]{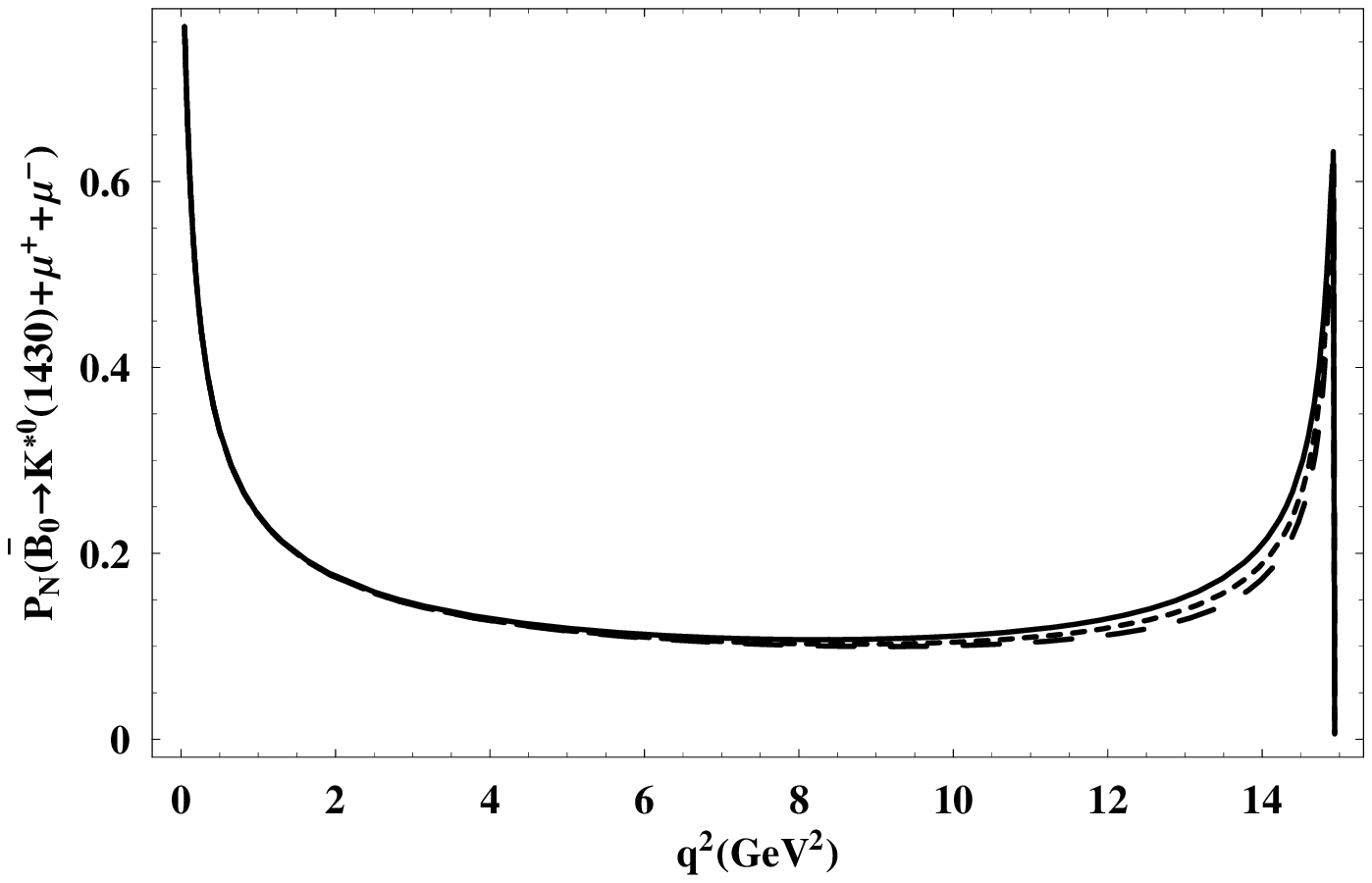} \put (-350,220){(a)} \put
(-100,220){(b)} &
\end{tabular}%
\end{center}
\caption{The dependence of Longitudinal (Fig. 3a) and Normal lepton
polarization (Fig. 3b) of $\bar{B}_{0}\rightarrow
K_{0}^{\ast }(1430)\protect\mu^{+}\protect\mu^{-}$ on $q^2$ for different
values of the input parameters. Solid value corresponds to the central value, dotted line is for maximum
value and long dashed line is for minimum value of input parameters.}
\label{Longitudinal-Normal polarization muons}
\end{figure}
\begin{figure}[h]
\begin{center}
\begin{tabular}{cc}
\vspace{-2cm} \includegraphics[scale=0.6]{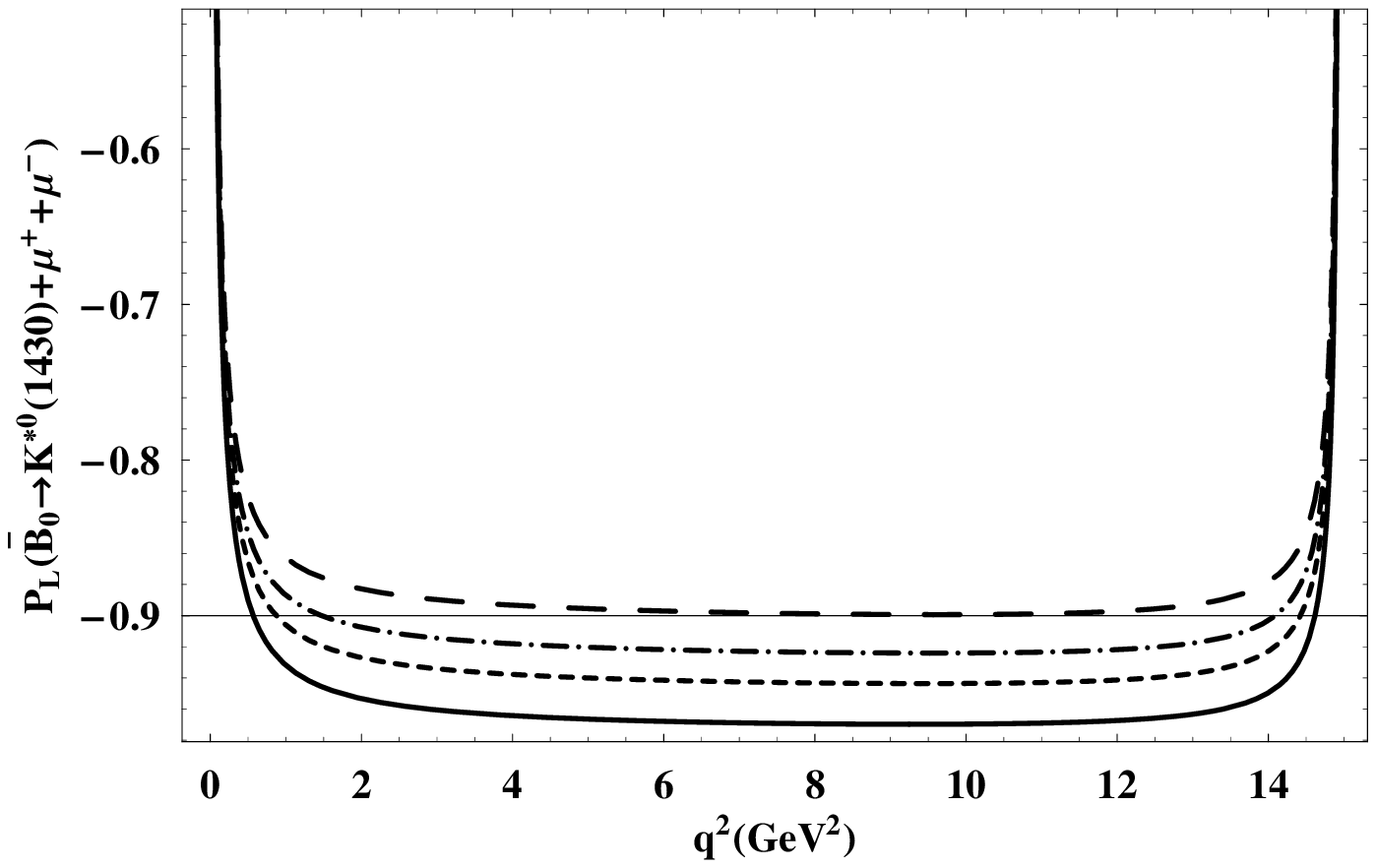} %
\includegraphics[scale=0.6]{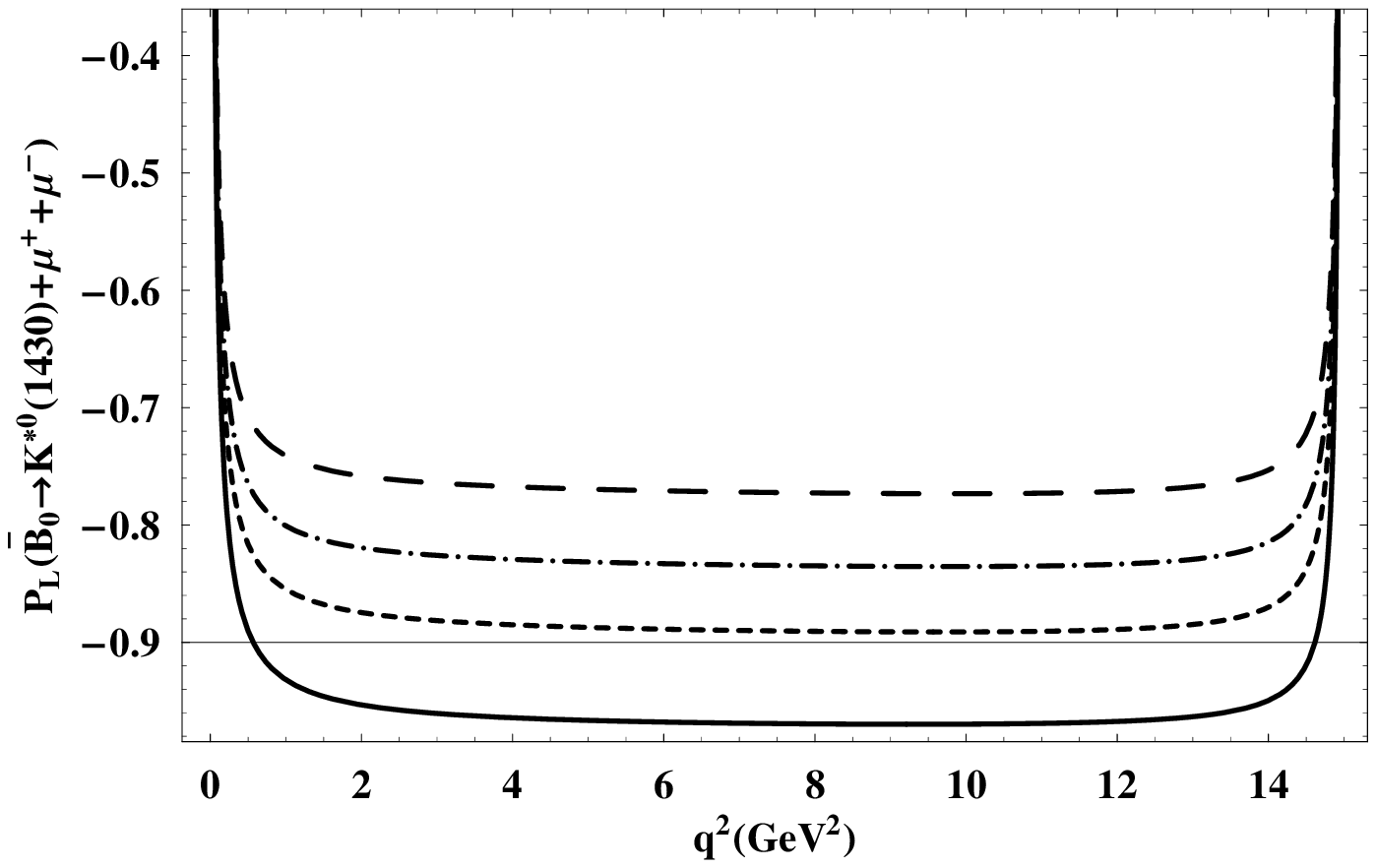} \put (-350,220){(a)} \put
(-100,220){(b)} &  \\
\includegraphics[scale=0.6]{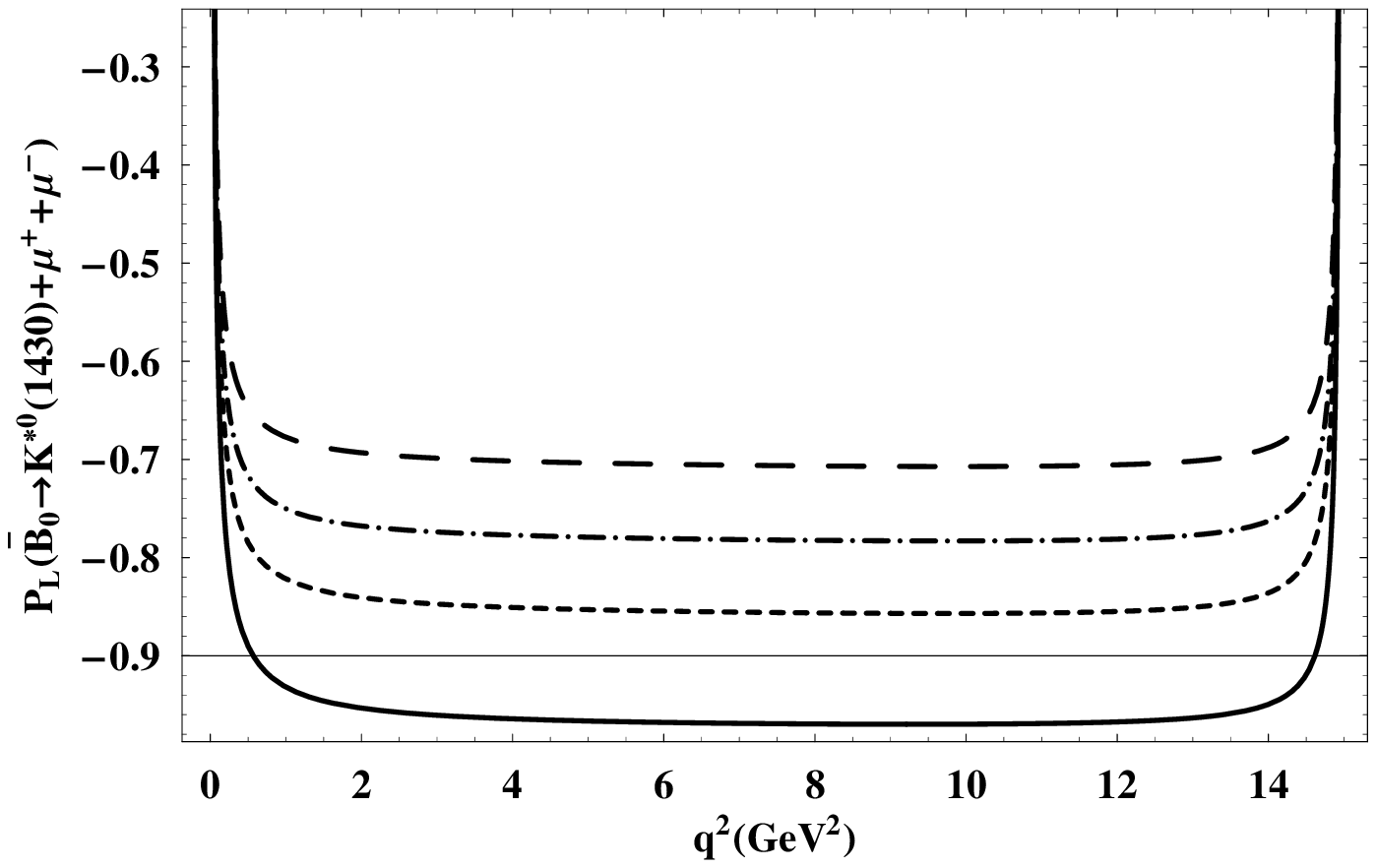} \includegraphics[scale=0.6]{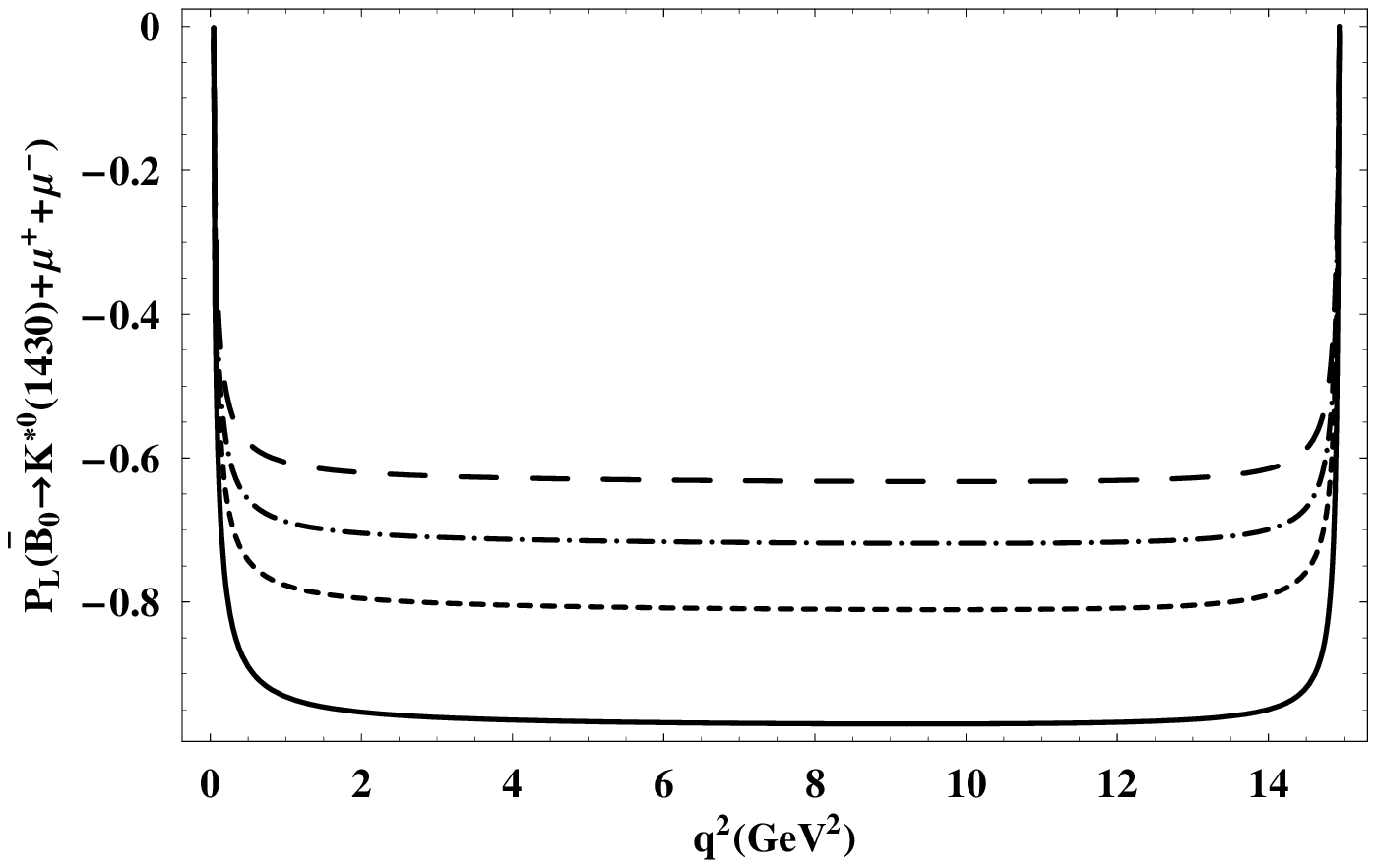}
\put (-350,220){(c)} \put (-100,220){(d)} &
\end{tabular}%
\end{center}
\caption{The dependence of Longitudinal lepton polarization of $\bar{B}_{0}\rightarrow
K_{0}^{\ast }(1430)\protect\mu ^{+}\protect\mu ^{-}$ on $q^2$ for different
values of $m_{t^{\prime }}$ and $\left\vert V_{t^{\prime }b}^{\ast
}V_{t^{\prime }s}\right\vert $. The values of fourth generation parameters
and the legends are same as in Fig.1}
\label{Longitudinal polarization muons}
\end{figure}
\begin{figure}[h]
\begin{center}
\begin{tabular}{cc}
\vspace{-2cm} \includegraphics[scale=0.6]{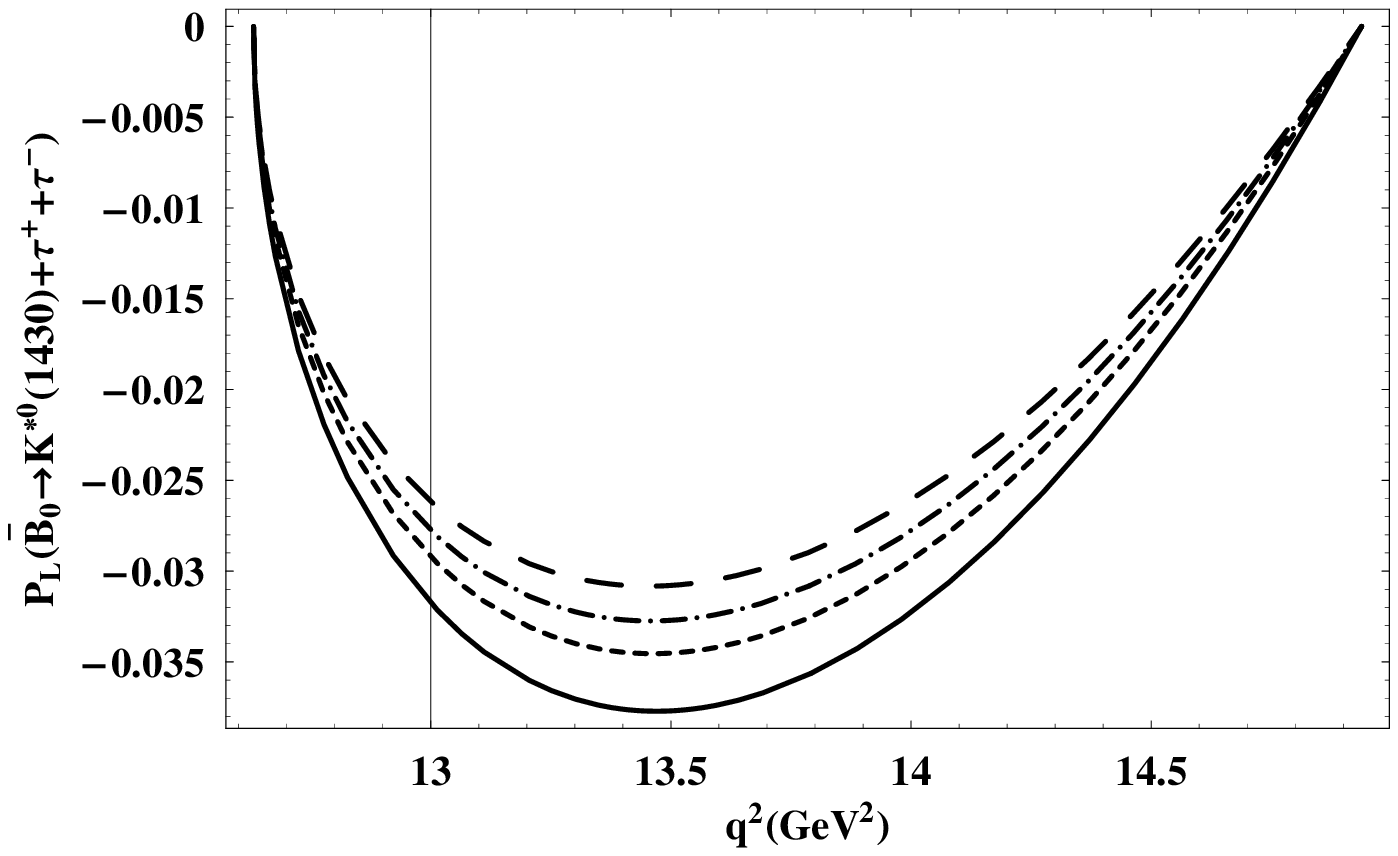} %
\includegraphics[scale=0.6]{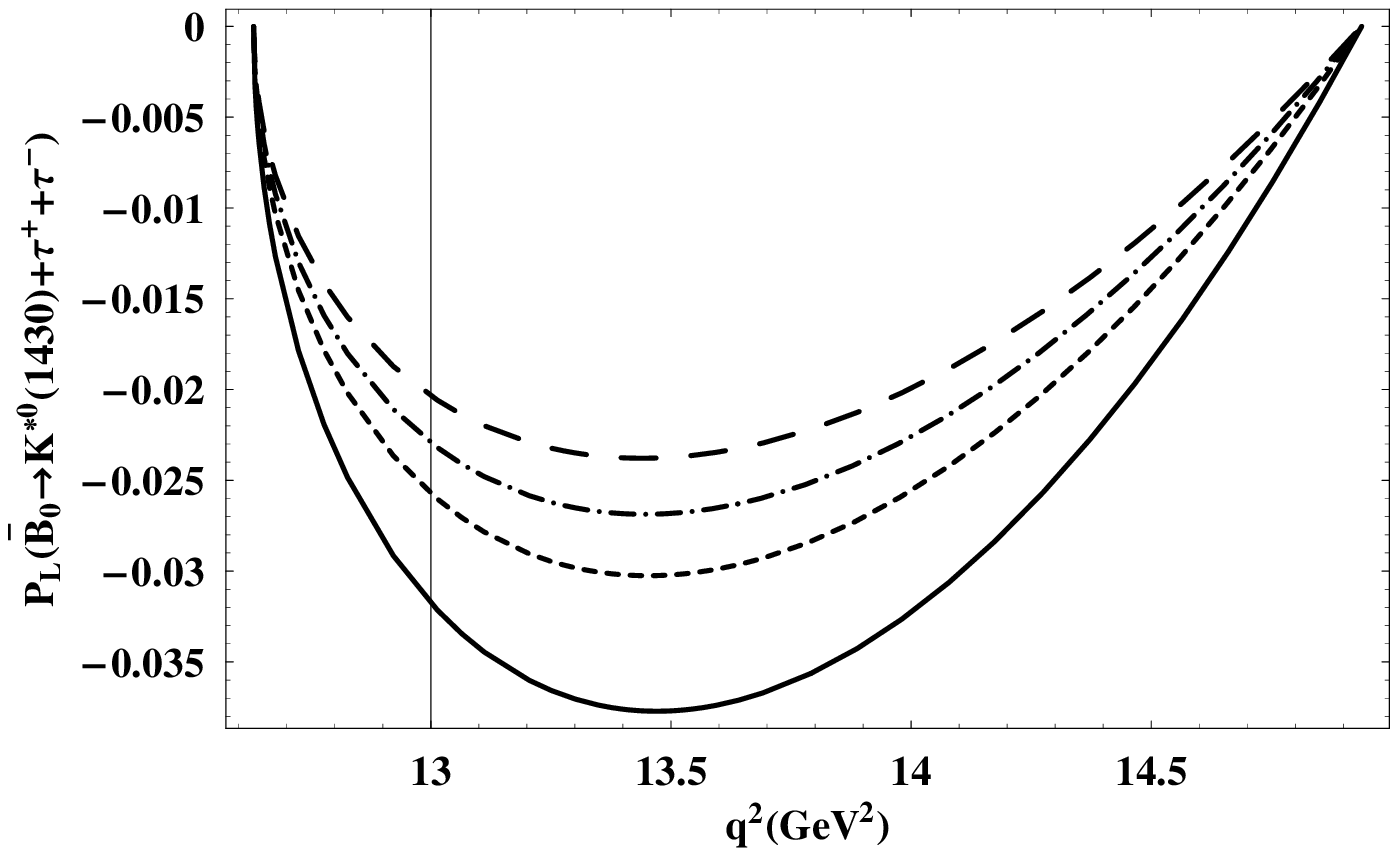} \put (-350,220){(a)} \put
(-100,220){(b)} &  \\
\includegraphics[scale=0.6]{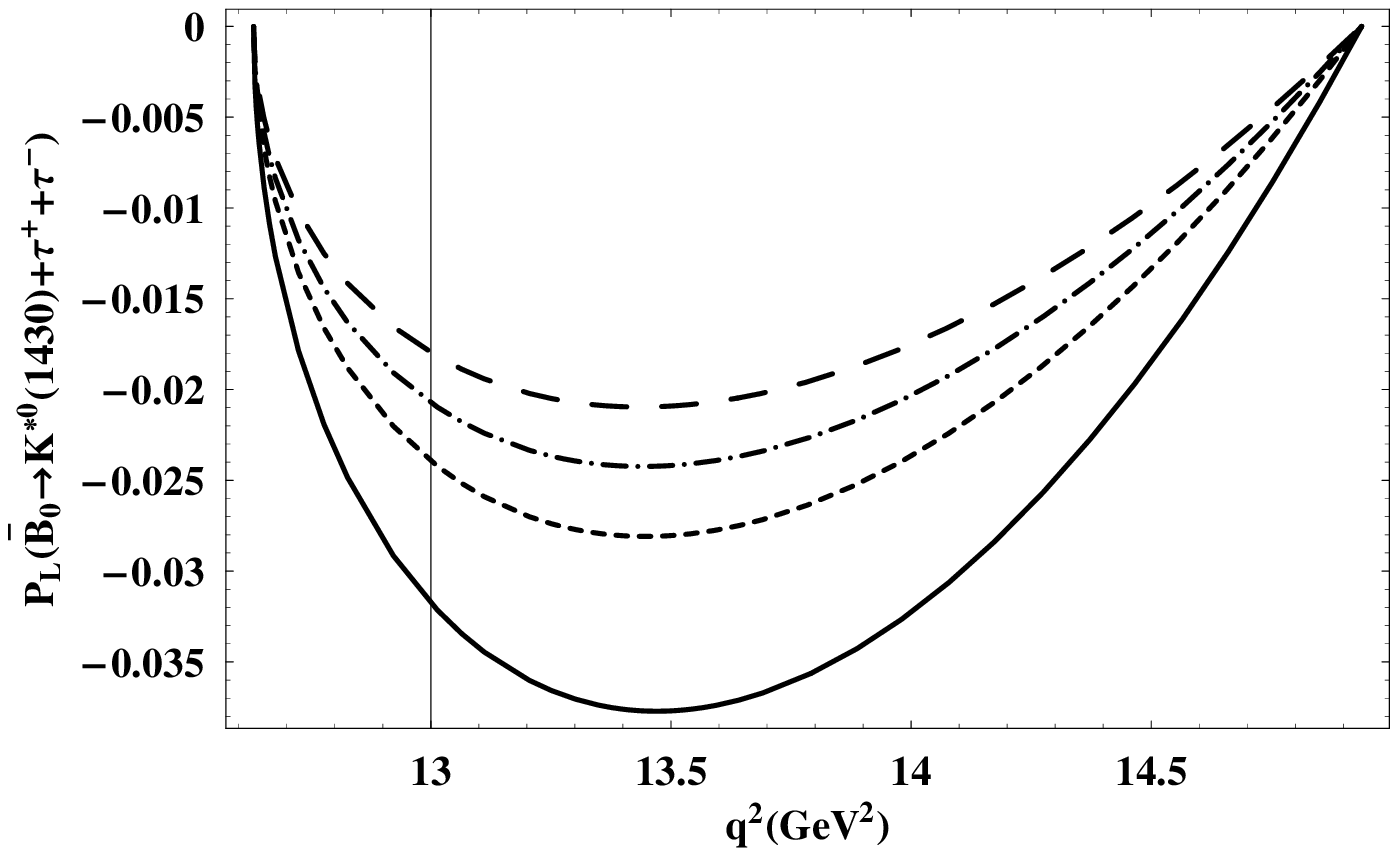} \includegraphics[scale=0.6]{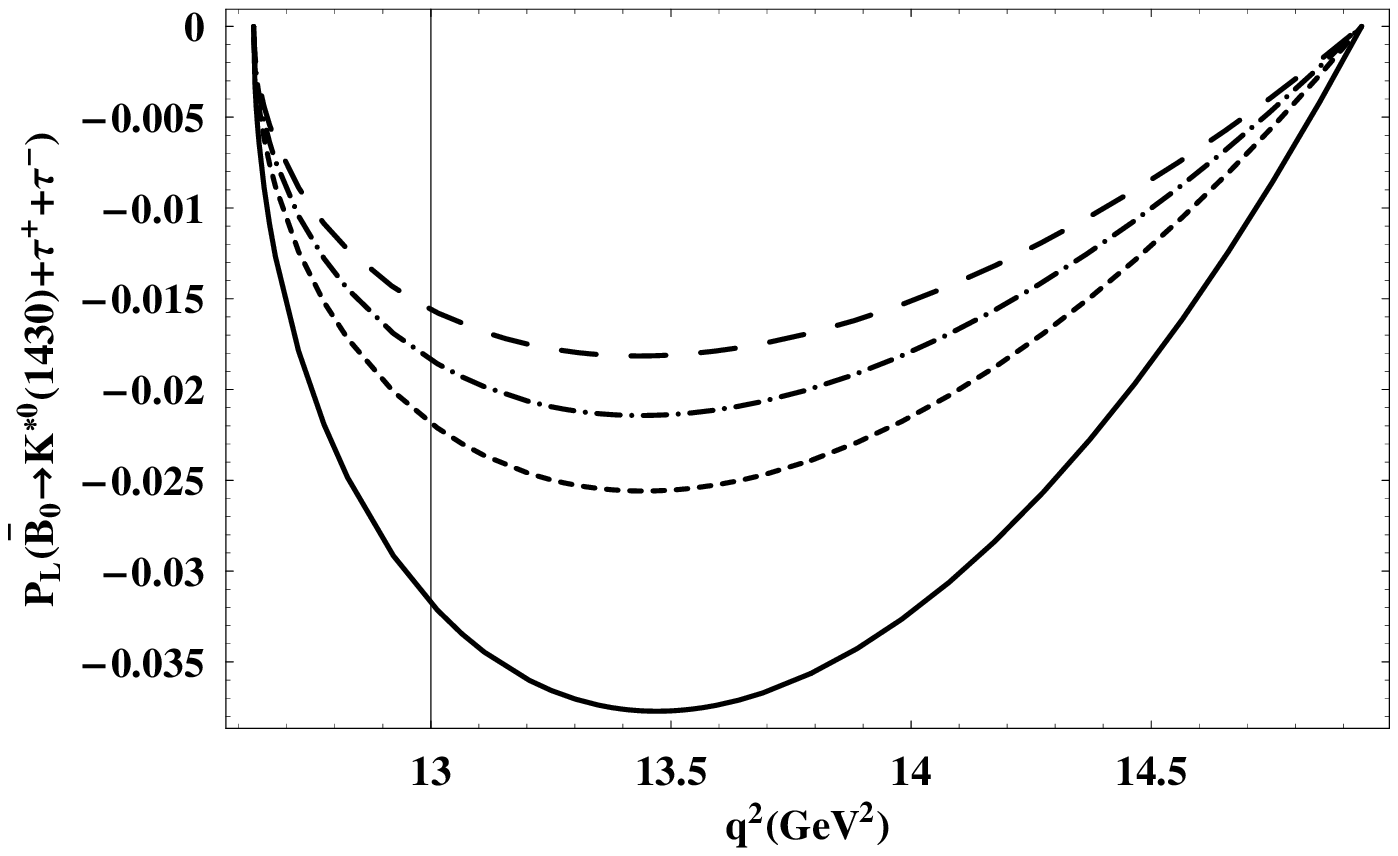}
\put (-350,220){(c)} \put (-100,220){(d)} &
\end{tabular}%
\end{center}
\caption{The dependence of Longitudinal lepton polarization of $\bar{B}_{0}\rightarrow
K_{0}^{\ast }(1430)\protect\tau ^{+}\protect\tau ^{-}$ on $q^2$ for different
values of $m_{t^{\prime }}$ and $\left\vert V_{t^{\prime }b}^{\ast
}V_{t^{\prime }s}\right\vert $. The values of fourth generation parameters
and the legends are same as in Fig.1}
\label{Longitudinal polarization tauons}
\end{figure}

The numerical results for the decay rates and polarization asymmetries of
the lepton are presented in Figs. 1-7. Figs. 1 and 2 describes the
differential decay rate of $B\rightarrow K^{\ast }_0(1430)l^{+}l^{-}$, from
which one can see that the fourth generation effects are quite distinctive
from that of the SM3 both in the small and large momentum transfer region.
At small value of $s$ the dominant contribution comes from $C_{7}^{tot}$
where as at the large value of $q^{2}$ the major contribution is from the $Z$
exchange i.e $C_{10}^{tot}$ which is sensitive to the mass of the fourth
generation quark $m_{t^{\prime }}$. Furthermore, for both the channels, the
branching ratios are enhanced sizably interms of $m_{t^{\prime }}$ and $%
\left\vert V_{t^{\prime }b}^{\ast }V_{t^{\prime }s}\right\vert $ and for $%
m_{t^{\prime }}=600$ and $\left\vert V_{t^{\prime }b}^{\ast }V_{t^{\prime
}s}\right\vert =1.4\times 10^{-2}$ the branching ratios are increased by an
order of magnitude.

As an exclusive decay, there are different source of
uncertainties involved in the calculation of the above said
decay. The major uncertainties in the numerical analysis of
$\bar{B^0}\to K^{*}_0(1430)l^{+}l^{-}$ decay originated from the
$\bar{B^0}\to K^{*}_0(1430)$ transition form factors calculated in the LCSR
approach as shown in Table I, which can bring about
almost $40\%$ errors to the differential decay rate of above mentioned decay,
which showed that it is not a very suitable tool to
look for the new physics. The large uncertainties involved
in the form factors are mainly from the variations of the
decay constant of $K^{*}_0(1430)$ meson and the Gengenbauer
moments in its distribution amplitudes. There are also
some uncertainties from the strange quark mass $m_s$, which
are expected to be very tiny on account of the negligible
role of $m_s$ suppressed by the much larger energy scale of
$m_{b}$. Moreover, the uncertainties of the charm quark and
bottom quark mass are at the $1\%$ level, which will not play
significant role in the numerical analysis and can be
dropped out safely. It also needs to be stressed that these
hadronic uncertainties almost have no influence on the
various asymmetries including the lepton polarization
asymmetry on account of the serious cancelation among
different polarization states and this make them the best
tool to look for physics beyond the SM. This has already been described in
ref.\cite{YuMing} and was shown in Fig. 3(a,b) for the longitudinal and normal lepton
polarization asymmetries.
\begin{figure}[h]
\begin{center}
\begin{tabular}{cc}
\vspace{-2cm} \includegraphics[scale=0.6]{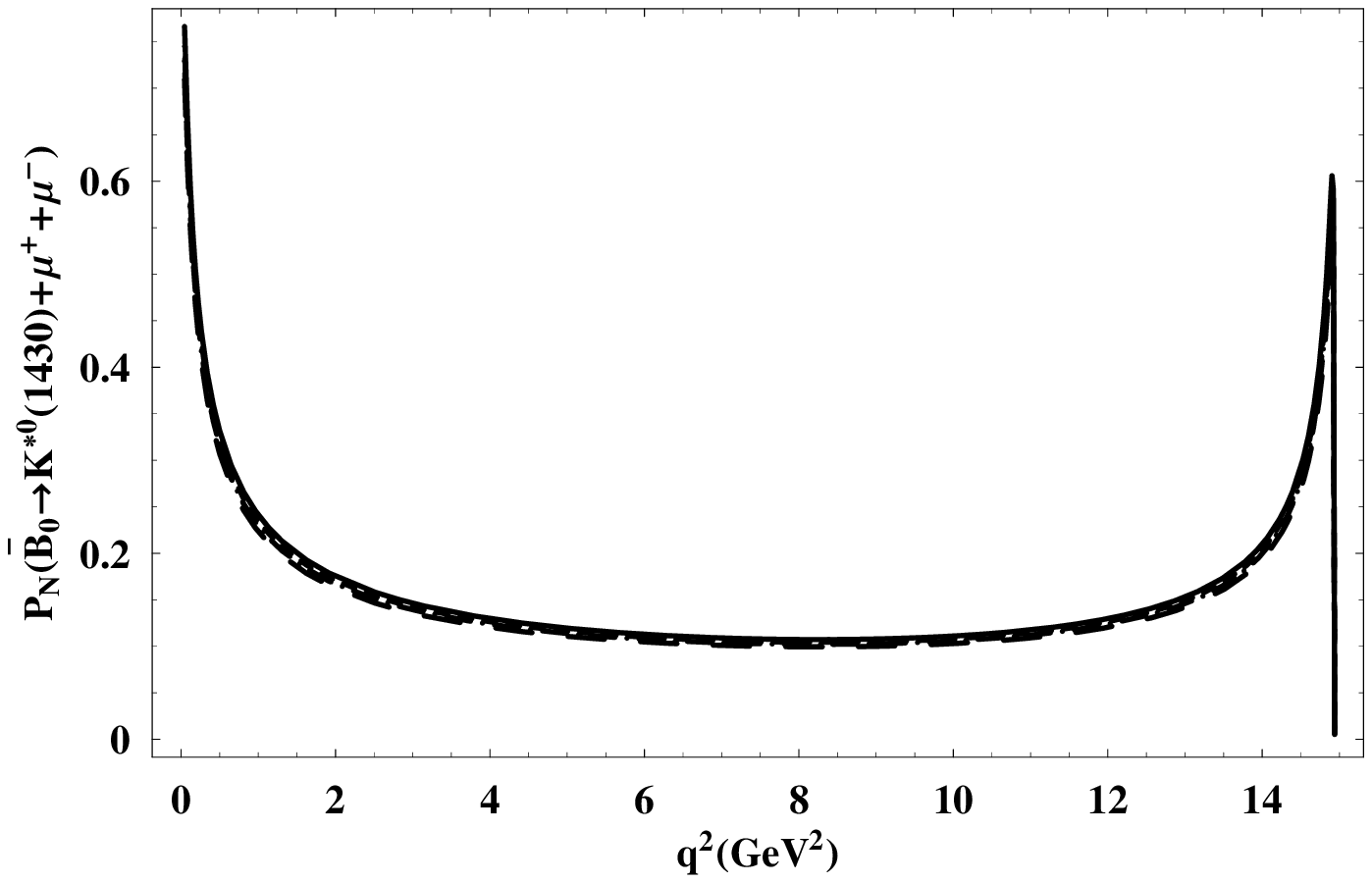} %
\includegraphics[scale=0.6]{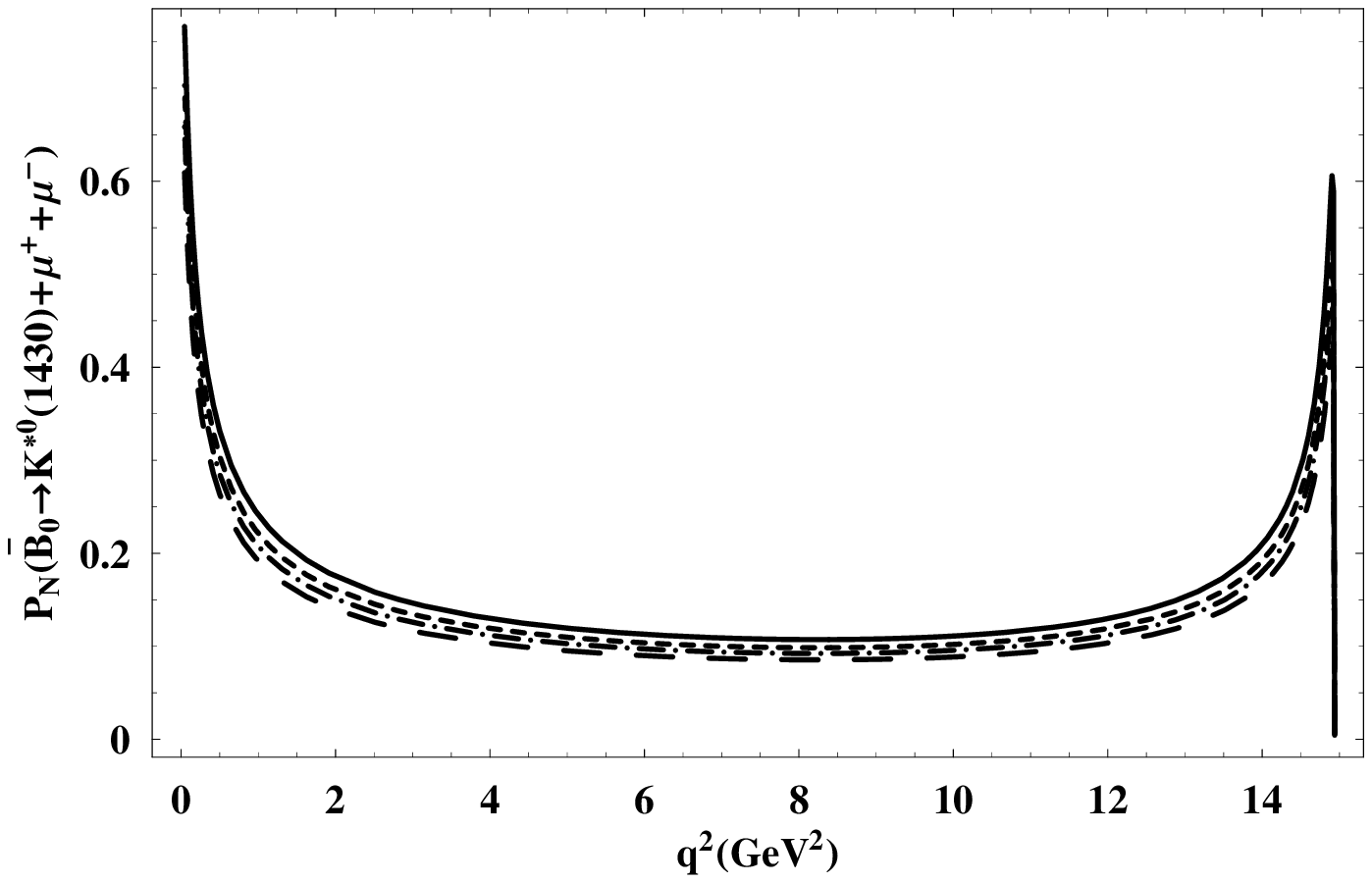} \put (-350,220){(a)} \put
(-100,220){(b)} &  \\
\includegraphics[scale=0.6]{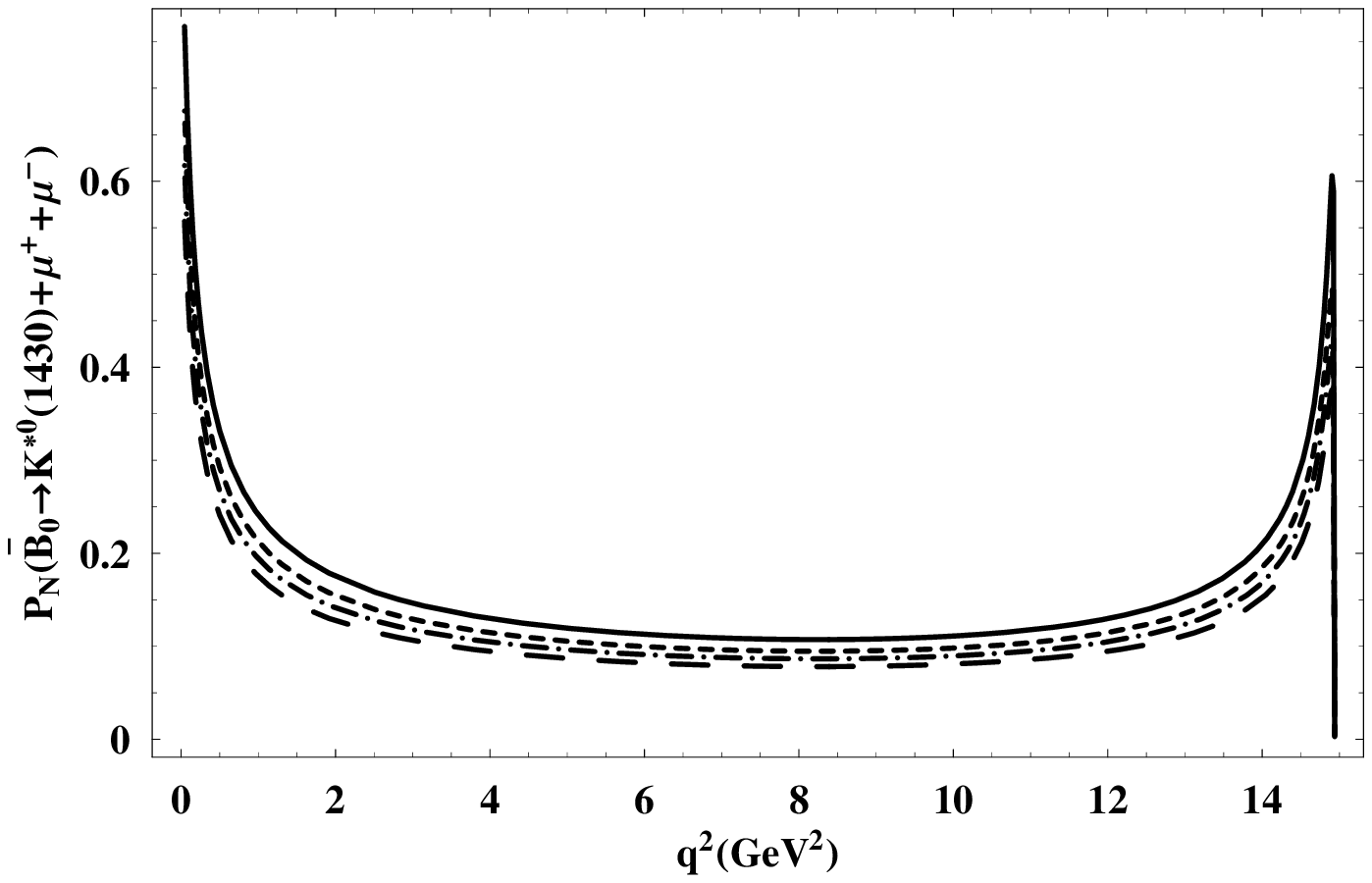} \includegraphics[scale=0.6]{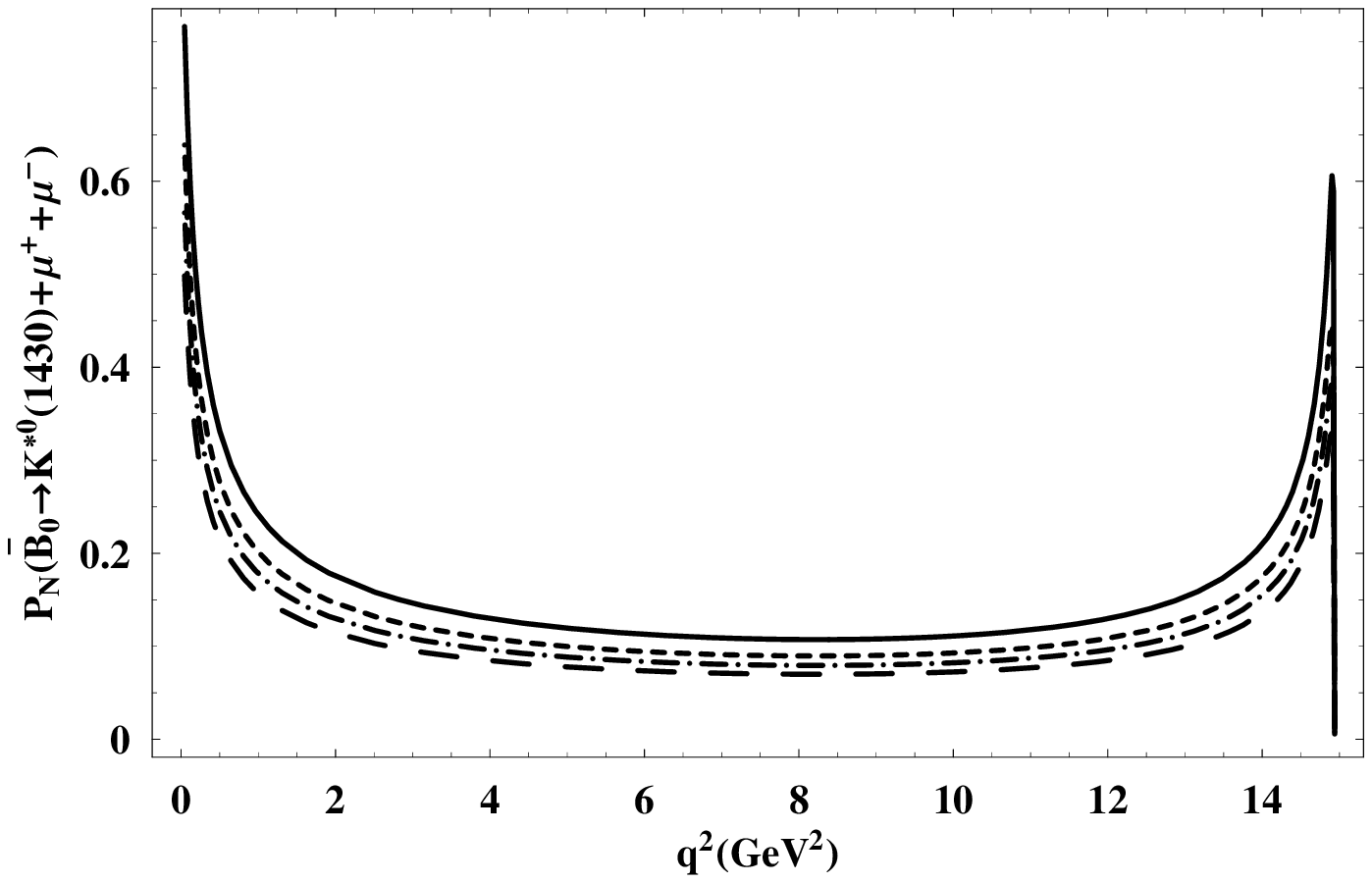}
\put (-350,220){(c)} \put (-100,220){(d)} &
\end{tabular}%
\end{center}
\caption{The dependence of Normal lepton polarization of $\bar{B}_{0}\rightarrow
K_{0}^{\ast }(1430)\protect\mu^{+}\protect\mu^{-}$ on $q^2$ for different
values of $m_{t^{\prime }}$ and $\left\vert V_{t^{\prime }b}^{\ast
}V_{t^{\prime }s}\right\vert $. The values of fourth generation parameters
and the legends are same as in Fig.1}
\label{Normal polarization muons}
\end{figure}
\begin{figure}[h]
\begin{center}
\begin{tabular}{cc}
\vspace{-2cm} \includegraphics[scale=0.6]{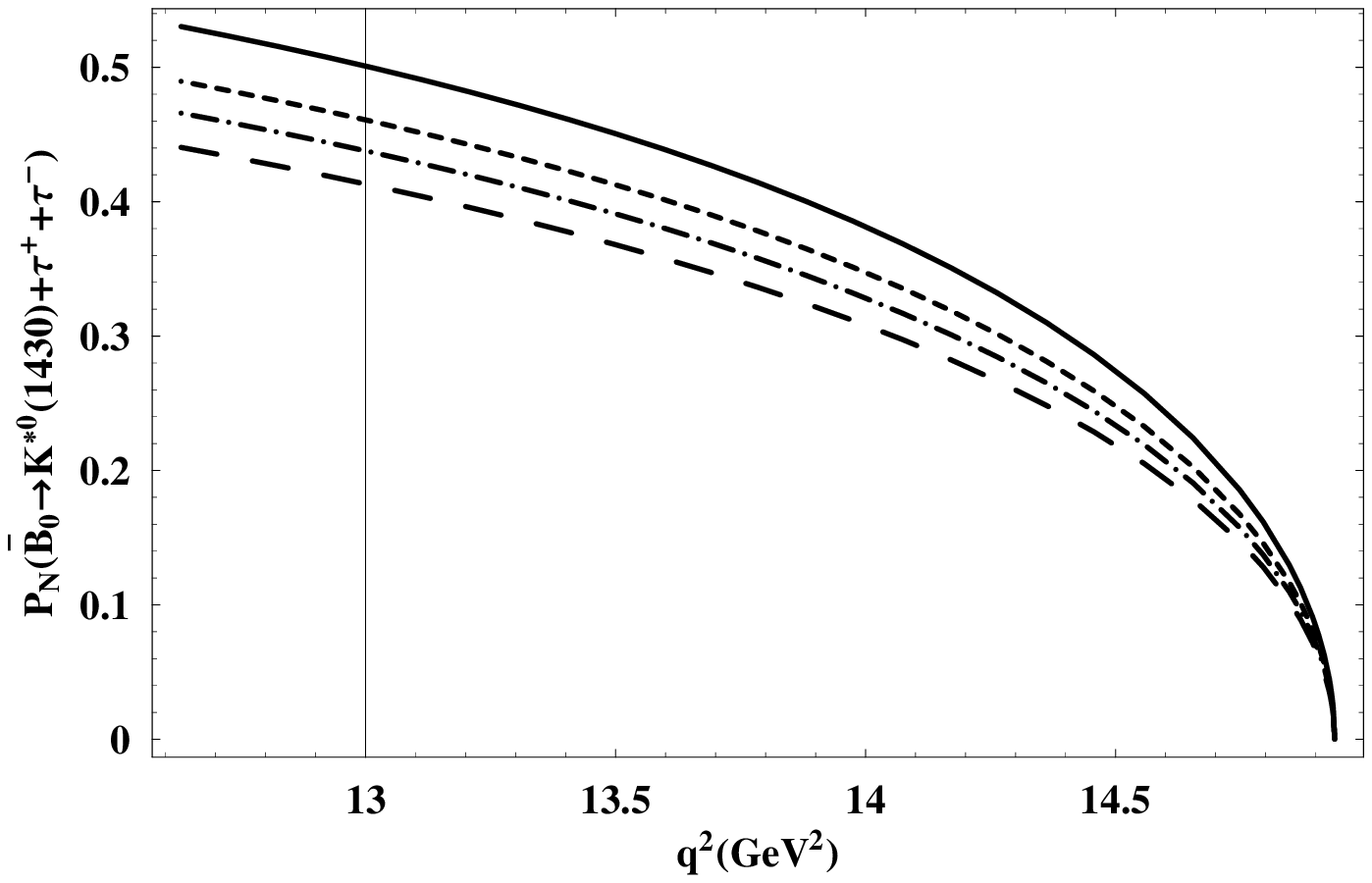} %
\includegraphics[scale=0.6]{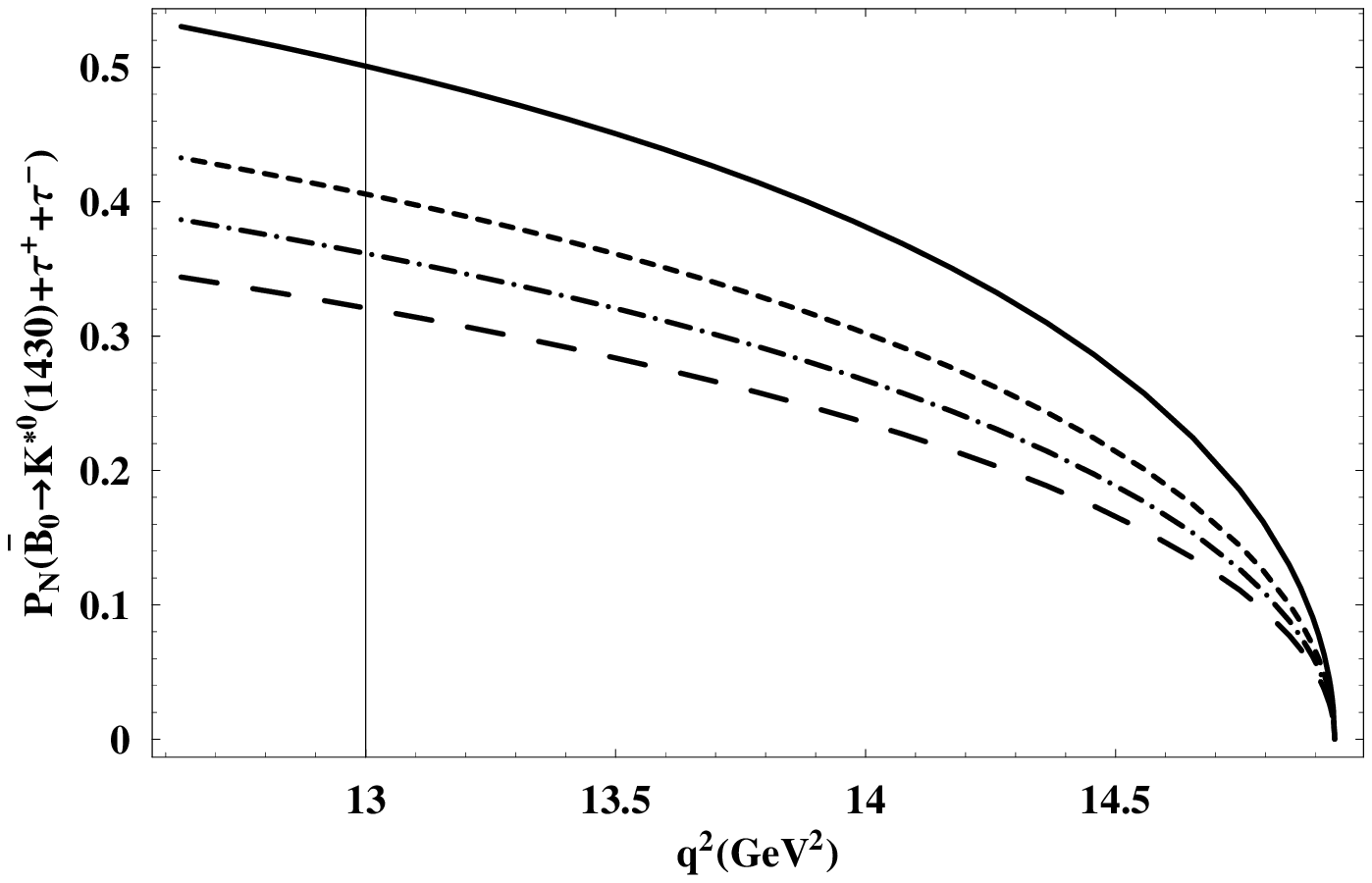} \put (-350,220){(a)} \put
(-100,220){(b)} &  \\
\includegraphics[scale=0.6]{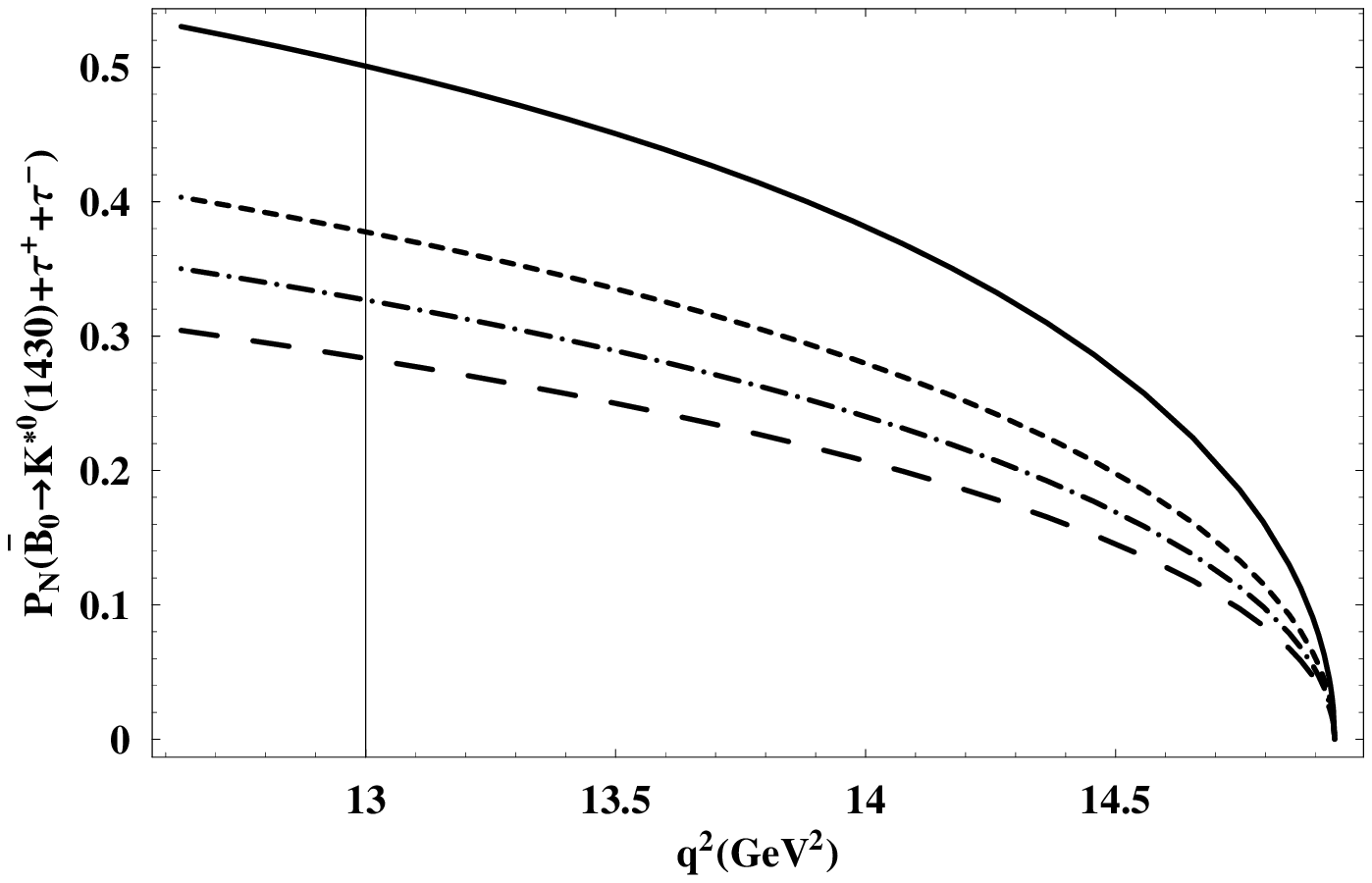} \includegraphics[scale=0.6]{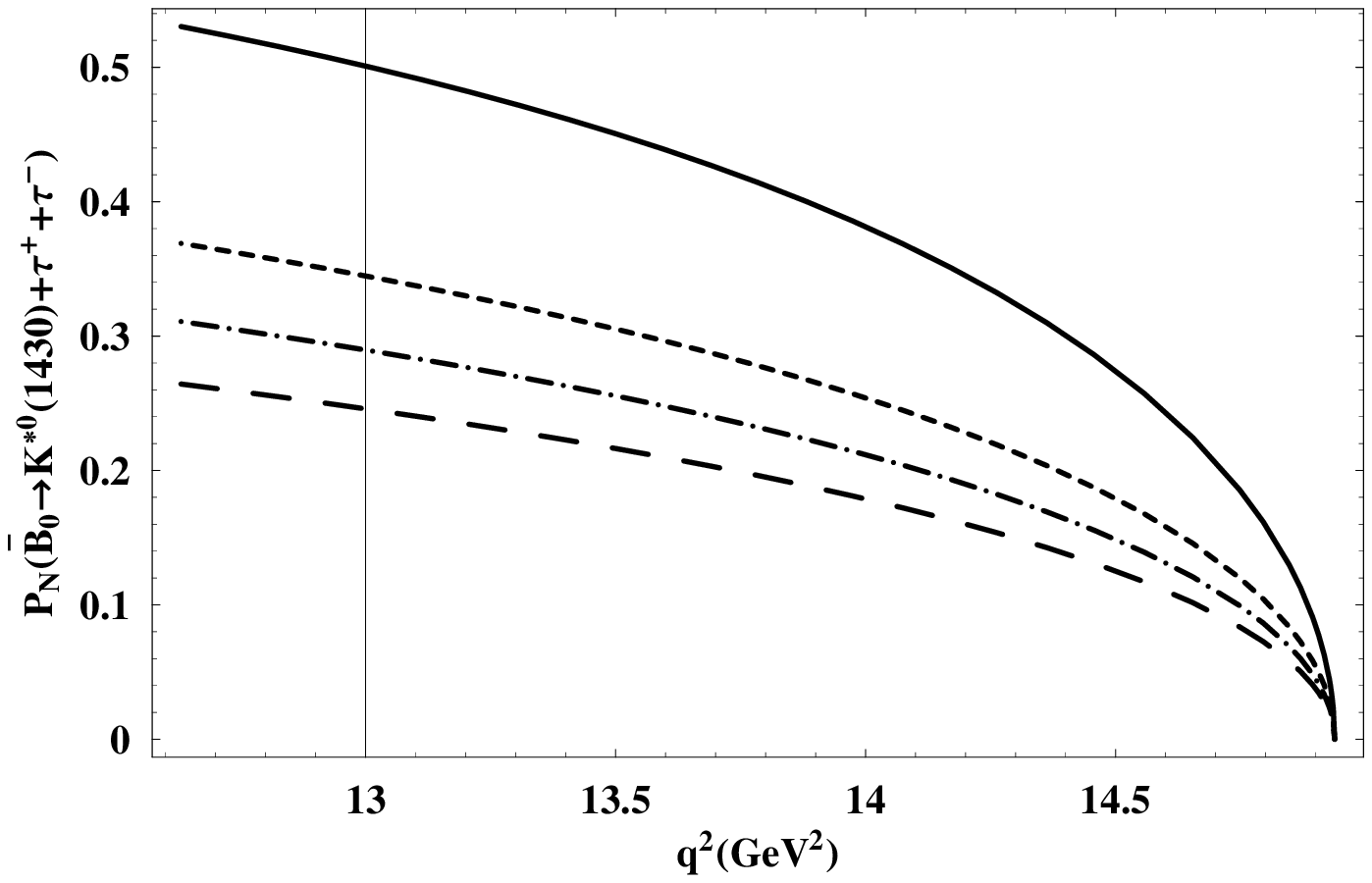}
\put (-350,220){(c)} \put (-100,220){(d)} &
\end{tabular}%
\end{center}
\caption{The dependence of Normal lepton polarization of $\bar{B}_{0}\rightarrow
K_{0}^{\ast }(1430)\protect\tau^{+}\protect\tau^{-}$ on $q^2$ for different
values of $m_{t^{\prime }}$ and $\left\vert V_{t^{\prime }b}^{\ast
}V_{t^{\prime }s}\right\vert $. The values of fourth generation parameters
and the legends are same as in Fig.1}
\label{Normal polarization tauons}
\end{figure}

Fig. 4(a,b,c,d) shows the dependence of longitudinal lepton polarization
asymmetry for the $B\rightarrow K_{0}^{\ast }\l ^{+}\l ^{-}$ decay on the
square of momentum transfer for different values of $m_{t^{\prime }}$ and $%
\left\vert V_{t^{\prime }b}^{\ast }V_{t^{\prime }s}\right\vert $. The value
of longitudinal lepton polarization for muon is around $-1$ in the SM3 and
we have significant deviation in this value in SM4. Just in the case of $%
m_{t^{\prime }}=600$ and $\left\vert V_{t^{\prime }b}^{\ast }V_{t^{\prime
}s}\right\vert =1.4\times 10^{-2}$ the value of the longitudinal lepton
polarization becomes $-0.6$ which will help us to see experimentally the SM4
effects in these flavor decays. In large $q^{2}$ region, the longitudinal
lepton polarization approaches to zero both in the SM3 and SM4 which is due
to the factor $\lambda (m_{B}^{2},m_{K_{0}^{\ast }}^{2},q^{2})$ that
approaches to zero at large value of $q^{2}$. Similar effects can been seen
for the final state tauon (c.f. Fig. 5) but the value for this case is too
small to measure experimentally.

The dependence of normal lepton polarization asymmetries for $B\rightarrow
K_{0}^{\ast }l^{+}l^{-}$ on the momentum transfer square are presented in
Figs. 6 and 7. In terms of Eq. (\ref{Normal-polarization}), one can see that
it is proportional to the mass of the final state lepton and for $\mu $ its
values is expected to be small and Fig. 6(a,b,c,d) displays it in the SM3 as
well as SM4 for the different value of fourth generation parameters. In SM4,
one can see a slight shift from the SM3 value which, however, is too small
to measure experimentally. Now, for the $\tau ^{+}\tau ^{-}$ channel, Eq. (%
\ref{Normal-polarization}) we will have a large value of normal lepton
polarization compared to the $\mu ^{+}\mu ^{-}$ case in the SM3. Fig. 7
shows that there is a significant decrease in the value of $P_{N}$ in SM4
compared to SM3 and its experimental measurement will give us some clue
about the fourth generation of quarks.

Now, from Eq. (\ref{expression-TPolarization}) we can see that it is
proportional to the lepton mass as well as to the form factor $f_{-}(q^{2})$
which is an order of magnitude smaller than the $f_{+}\left( q^{2}\right) $
and $f_{T}\left( q^{2}\right) $. This makes the transverse lepton
polarization asymmetry almost to be zero in the SM3 as well as in the SM4
and it is non-zero only in the models where we have new operators, e.g.
scalar type operators in the MSSM.

\section{Conclusions}

We have carried out the study of invariant mass spectrum and polarization
asymmetries of semileptonic decays $\bar{B}_{0}\rightarrow K^{\ast}_0(1430)
l^{+}l^{-} $ ($l=\mu ,\tau $) decays in SM4. Particularly, we analyzed the
sensitivity of these physical observables on the fourth generation quark
mass $m_{t\prime}$ as well as the the mixing angle $\left\vert V_{t^{\prime
}b}^{\ast }V_{t^{\prime }s}\right\vert $ and the main outcomes of this study
can be summarized as follows:

\begin{itemize}
\item The differential decay rates deviate sizably from that of the SM
especially both in the small and large momentum transfer region. These
effects are significant and the branching ration increases by an order of
magnitude for $m_{t\prime}$ $=$ $600$ GeV and $\left\vert V_{t^{\prime
}b}^{\ast }V_{t^{\prime }s}\right\vert $ $=$ $1.4\times 10^{-2}$.

\item It has been shown in the literature \cite{YuMing} that the value of the forward-backward
asymmetry for $\bar{B}_{0} \to K^{\ast}_0(1430) l^{+}l^{-}
$ is non zero only in the models where we have the scalar type operators (like SUSY models).
Now, due to the absence of scalar type operators in the SM3 as well as in
SM4 the forward-backward asymmetry for the decay $\bar{B}_{0} \to K^{\ast}_0(1430) l^{+}l^{-}
$ is zero in both these models.

\item The longitudinal, normal and transverse polarizations of leptons are
calculated in the SM4. It is found that the SM effects are very promising
which could be measured at future experiments and shed light on the new
physics beyond the SM. It is hoped that this can be measurable at future
experiments like LHC and BTeV machines where a large number of $\ b \bar {b}
$ pairs are expected to be produced.
\end{itemize}

In short, the experimental investigation of observables, like decay rates,
forward-backward asymmetry, lepton polarization asymmetries and the
polarization asymmetries of $\bar{B}_{0}\rightarrow K^{\ast}_0(1430) l^{+}l^{-}$ ($%
l=\mu ,\tau $) decay will be used to search for the SM4 effects and will
help us to put constraints on fourth generation parameters in an indirect
way.

\section*{Acknowledgements}

The author would like to thank Ali Paracha and Ishtiaq Ahmed for some useful
discussions. He would also like to thank the facilities provided by National
Centre for Physics during the completion of this work.


\begin{thebibliography}{99}
\bibitem{Cabibbo} N. Cabibbo, Phys. Rev. Lett. \textbf{10} (1963) 531

\bibitem{Kobayashi} M. Kobayashi and T. Maskawa, Prog. Theor. Phys. \textbf{%
49} (1973) 652.

\bibitem{Lunghi} E.Lunghi and A. Soni, JHEP 0709 (2007) 053.

\bibitem{Bona} M. Bona et al. [UT Fit Collaboration], arXiv: 0707.0636.

\bibitem{Lunghi1} E. Lunghi and A. Soni, arXiv: 0803.4340.

\bibitem{Frampton} P. H. Frampton, P. Q. Hung, and M. Sher, Phys. Rept.
\textbf{330} (2000) 263, hep-ph/9903387, B. Holdom et al., arXiv:
hep-ph/0904.4698.

\bibitem{Kribs} G. D. Kribs, T. Plehn, M. Spannowsky, and T. M. P. Tait,
Phys. Rev. \textbf{D76} (2007) 075016, 0706.3718 .

\bibitem{Hashimoto} M. Hashimoto, arXiv:1001.4335.

\bibitem{Alwall} J. Alwall et al., Eur. Phys. J. \textbf{C49} (2007)
791-801, hep-ph/0607115

\bibitem{Chanowitz} M. S. Chanowitz, Phys. Rev. \textbf{D79} (2009) 113008,
arXiv: 0904.3570 .

\bibitem{Novikov} V. A. Novikov, A. N. Rozanov, and M. I. Vysotsky, arXiv:
0904.4570.

\bibitem{Erler} J. Erler and P. Langacker, arXiv: 1003.3211.

\bibitem{Polonsky} H.-J. He, N. Polonsky, S. Su,
Phys. Rev. \textbf{D64} (2001) 053004, [hep-ph/0102144].

\bibitem{Hung} P. Q. Hung and C. Xiong, arXiv: 0911.3890.

\bibitem{Xiong} P. Q. Hung and C. Xiong, arXiv: 0911.3892.

\bibitem{Babu} K. S. Babu, X. G. He, X. Li, and S. Pakvasa, Phys. Lett.
\textbf{B205} (1988) 540.

\bibitem{London} D. London, Phys. Lett. \textbf{B234} (1990) 354.

\bibitem{Dincer} Y. Dincer, Phys. Lett. \textbf{B505} (2001) 89.

\bibitem{Arhrib} A. Arhrib and W.-S. Hou, Eur. Phys. J. \textbf{C27} (2003)
555-561, hep-ph/0211267 .

\bibitem{Hou1} W.-S. Hou, M. Nagashima, and A. Soddu, Phys. Rev. \textbf{D72}
(2005) 115007, hep-ph/0508237.

\bibitem{Hou2} W.-S. Hou, M. Nagashima, and A. Soddu, Phys. Rev. \textbf{D76}
(2007) 016004, hep-ph/0610385.

\bibitem{Alieve} T. M. Aliev, A. Ozpineci and M. Savci, Nucl. Phys. \textbf{%
B585} (2000) 275. T. M. Aliev, A. Ozpineci and M. Savci, Eur. Phys. J.
\textbf{C29} (2003) 265.

\bibitem{Bashiry} V. Bashiry and K. Azizi, JHEP 0707 (2007) 64; V. Bashiry
and F. Flahati, arXiv: 0707.3242; F. Zolfagharpour and V. Bashiry, arXiv:
0707.4337; V. Bashiry and M. Bayer, arXiv: 0903.2631.

\bibitem{soniAlok} A. Soni, A. K. Alok, A. Giri, R. Mohanta, and S. Nandi,
arXiv: 0807.1971.

\bibitem{Herrera} J. A. Herrera, R. H. Benavides, and W. A. Ponce, Phys.
Rev. \textbf{D78} (2008) 073008, 0810.3871 .

\bibitem{Bobrowski} M. Bobrowski, A. Lenz, J. Riedl, and J. Rohrwild, Phys.
Rev. \textbf{D79} (2009) 113006, 0902.4883 .

\bibitem{Eilam} G. Eilam, B. Melic, and J. Trampetic, Phys. Rev. \textbf{D80}
(2009) 116003, 0909.3227 .

\bibitem{Giri} A. Soni, A. K. Alok, A. Giri, R. Mohanta, and S. Nandi,
arXiv: 1002.0595.

\bibitem{Buras1} A. J. Buras, B. Duling, T. Feldmann, T. Heidsieck, C.
Promberger, and S. Recksiegel, arXiv: 1002.2126.

\bibitem{Hou3} W. S. Hou and C. Y. Ma, arXiv: 1004.2186.

\bibitem{lunghinew} E. Lunghi and A. Soni, arXiv: 1007.4015.

\bibitem{susy1} Z. Murdock, S. Nandi, and Z. Tavartkiladze, Phys. Lett.
\textbf{B668} (2008) 303-307, arXiv: 0806.2064 .

\bibitem{susy2} R. M. Godbole, S. K. Vempati, and A. Wingerter, arXiv:
0911.1882, and references therein.

\bibitem{Duling} A. J. Buras, B. Duling, T. Feldmann, T. Heidsieck, C.
Promberger and S. Recksiegel, arXiv: 1004.4565.

\bibitem{YuMing} Yu-Ming Wang, M. Jamil Aslam, Cai-Dian Lu, Phys. Rev.
\textbf{D78} (2008) 014006; M. Jamil Aslam, Cai-Dian Lu, Yu-Ming Wang,
Phys. Rev. \textbf{D79} (2009) 074007.

\bibitem{Buchalla} G. Buchalla, A. J. Buras and M. E. Lautenbacher, Rev.
Mod. Phys. \textbf{68} (1996) 1125.

\bibitem{Buras} A. J. Buras and M. Munz, Phys. Rev. \textbf{D52} (1995) 186;
A. J. Buras, M. Misiak, M. Munz and S. Pokorski, Nucl. Phys. \textbf{424}
374.

\bibitem{Kim} C.S. Kim, T. Morozumi, A.I. Sanda, Phys. Lett. B \textbf{218}
(1989) 343.

\bibitem{Ali} A. Ali, T. Mannel and T. Morozumi, Phys. Lett. \textbf{B273}
(1991) 505.

\bibitem{Kruger} F. Kruger and L. M. Sehgal, Phys. Lett. \textbf{380} (1996)
199.

\bibitem{Grinstein} B. Grinstein, M. J. Savag and M. B. Wise, Nucl. Phys.
\textbf{B319} (1989) 271.

\bibitem{Cella} G. Cella, G. Ricciardi adn A. Vicere, Phys. Lett. \textbf{%
B258} (1991) 212.

\bibitem{Bobeth} C. Bobeth, M. Misiak and J. Urban, Nucl. Phys. \textbf{B574}
(2000) 291.

\bibitem{Asatrian} H. H. Asatrian, H. M. Asatrian, C. Grueb and M. Walker,
Phys. Lett. B507 (2001) 162.

\bibitem{Misiak} M. Misiak, Nucl. Phys. \textbf{B393} (1993) 23, Erratum,
ibid. \textbf{B439} (1995) 461.

\bibitem{Huber} T. Huber, T. Hurth, E. Lunghi, arXiv: 0807.1940.

\bibitem{b to s 1}  D.~Melikhov, N.~Nikitin and S.~Simula,
Phys.\ Lett.\ B \textbf{430} (1998) 332 [arXiv:hep-ph/9803343].

\bibitem{b to s 2}  J.~M.~Soares, Nucl.\ Phys.\ B \textbf{367} (1991) 575.

\bibitem{b to s 3}  G.~M.~Asatrian and A.~Ioannisian,
Phys.\ Rev.\ D \textbf{54} (1996) 5642 [arXiv:hep-ph/9603318].

\bibitem{NF charm loop}  J.~M.~Soares,
Phys.\ Rev.\ D \textbf{53} (1996) 241 [arXiv:hep-ph/9503285].

\bibitem{c.q. geng 4}  C.~H.~Chen and C.~Q.~Geng,
Phys.\ Rev.\ D \textbf{64} (2001) 074001  [arXiv:hep-ph/0106193].

\bibitem{PDG} C.~Amsler \textit{et al.} [Particle Data Group],
Phys.\ Lett.\ B \textbf{667}, 1 (2008).

\bibitem{Aliev UED} T. M. Aliev and M. Savci, Eur. Phys. J. C \textbf{50 }%
(2007)\textbf{\ }91 [arXiv. hep-ph/0606225].
\end{thebibliography}
\end{document}